# DATA-DRIVEN FRAMEWORK DEVELOPMENT FOR PUBLIC SPACE QUALITY ASSESSMENT

**Sherzod Turaev[1] and Mary John[1]**

[1]Department of Computer Science and Software Engineering, College of Information Technology, United Arab Emirates University, Al Ain, United Arab Emirates.

Corresponding authors: Sherzod Turaev (e-mail: sherzod@uaeu.ac.ae), Mary John (e-mail: maryjohn@uaeu.ac.ae).

The authors thank the United Arab Emirates University for supporting this work through the UAEU Strategic Research Grant (G00003676) and the Abu Dhabi International Virtual Research Institute for Food Security in the Drylands (G00004017 - VRI-FS 120-21).

## ABSTRACT

Public space quality assessment lacks systematic methodologies that integrate factors across diverse spatial typologies while maintaining context-specific relevance. Current approaches remain fragmented within disciplinary boundaries, limiting comprehensive evaluation and comparative analysis across different space types. This study develops a systematic, data-driven framework for assessing public space quality through the algorithmic integration of empirical research findings. Using a 7-phase methodology, we transform 1,207 quality factors extracted from 157 peer-reviewed studies into a validated hierarchical taxonomy spanning six public space typologies: urban spaces, open spaces, green spaces, parks and waterfronts, streets and squares, and public facilities. The methodology combines semantic analysis, cross-typology distribution analysis, and domain knowledge integration to address terminological variations and functional relationships across space types. The resulting framework organizes 1,029 unique quality factors across 14 main categories and 66 subcategories, identifying 278 universal factors applicable across all space types, 397 space-specific factors unique to particular typologies, and 124 cross-cutting factors serving multiple functions. Framework validation demonstrates systematic consistency in factor organization and theoretical alignment with established research on public spaces. This research provides a systematic methodology for transforming empirical public space research into practical assessment frameworks, supporting evidence-based policy development, design quality evaluation, and comparative analysis across diverse urban contexts.

**KEYWORDS: Public** Space Quality, Taxonomic Framework, Space Typology, Quality Assessment, Hierarchical Classification.

# 1. INTRODUCTION

Public space quality assessment operates through fragmented methodological approaches. Urban planners often measure accessibility, environmental researchers study thermal comfort, and social scientists examine inclusivity. Each discipline tends to develop specialized tools that address only fragments of what makes public spaces function. Cities frequently invest resources in public space improvements without standardized methods to evaluate outcomes or compare effectiveness across different types of spaces. This methodological gap appears to limit evidence-based decision-making in urban planning and suggests barriers to systematic understanding of what creates successful public spaces [1, 2].

Different types of public spaces serve distinct functions within the urban fabric. Streets and sidewalks facilitate movement while occasionally hosting markets or festivals. Parks and green spaces provide natural experiences and recreational outlets. Civic plazas create formal gathering areas for cultural events and demonstrations. Indoor public facilities, such as libraries and community centers, offer specialized services and protection from the weather. Each type requires different design approaches and quality considerations based on its intended functions and surrounding context [3].

Efforts to assess the quality of public space have evolved over recent decades. Early approaches often emphasize quantitative measures such as area per capita or distance to nearest green space. Contemporary methods tend to incorporate qualitative dimensions that extend beyond these foundational metrics. Various specialized tools have emerged, such as walkability indices for streets [4], thermal comfort assessments for open areas [5], and inclusion metrics for playgrounds [6]. Some researchers focus on user perceptions through surveys and interviews, while others employ objective measurements of environmental conditions or behavioral observations. These assessment approaches, while valuable, often remain isolated within their respective disciplines. Urban designers may concentrate on spatial configuration and visual qualities, social scientists on usage patterns and inclusivity, environmental researchers on ecological functions and microclimates, and economists on property values and business impacts. This fragmentation creates challenges for holistic evaluation across different public space types and contexts [7, 8]. Our recent comprehensive narrative review [9] systematically analyzed quality factors across 157 studies spanning multiple public space typologies, revealing both universal principles and space-specific requirements. However, this empirical synthesis, while identifying critical patterns across space types, did not provide the systematic methodological framework needed to transform these findings into practical hierarchical assessment tools.

Current assessment frameworks appear to reflect this disciplinary fragmentation. Specialized tools tend to perform well within narrow domains but often prove less effective when applied across different space types. Walkability indices function adequately for streets, yet provide limited insight for park evaluation. Green space assessment methods frequently overlook social programming factors that appear important for urban plazas. Comfort measurement protocols developed for outdoor spaces may be inadequate for indoor public facilities. Each framework typically addresses specific quality dimensions while overlooking others, making a comprehensive evaluation challenging.

Moreover, existing frameworks typically emphasize certain space types while overlooking others. Park evaluation methods may not transfer effectively to street assessments, and criteria for urban squares may not adequately capture the qualities of natural green spaces. This typological specialization, while offering precision for specific contexts, limits comparative analysis and comprehensive urban planning [10]. As cities worldwide face challenges of growth, densification, climate change, and social inequality, the need for integrated assessments of public space quality becomes increasingly urgent. Municipal governments require evidence-based tools to prioritize investments. Designers require guidance to create responsive and inclusive environments. Communities deserve clear standards to advocate for neighborhood improvements. A multidimensional framework that acknowledges both universal and typology-specific quality factors could support better decision-making across these stakeholder groups [11]. This research examines three questions: How can quality factors from diverse public space studies be systematically organized into coherent frameworks? What relationships exist between universal quality requirements and space-specific characteristics? How can computational methods transform fragmented research findings into validated frameworks for practical applications?

To address these questions and limitations, the present study develops a domain-informed, data-driven framework that systematically integrates quality factors across major public space typologies. This approach builds upon previous narrative review findings and suggests a pathway toward more comprehensive assessment methods [9]. Through a comprehensive analysis of empirical research spanning parks, waterfronts, streets, squares, urban spaces, green spaces, open spaces, and public facilities, this study creates a unified hierarchical framework that balances universal principles with typology-specific requirements.

The 7-phase methodology employs a systematic approach to processing empirical research findings. The approach transforms descriptive factor identification into organized taxonomic structures. Semantic analysis resolves terminology inconsistencies across studies. Distribution analysis distinguishes universal requirements from space-specific needs. Hierarchical clustering organizes factors into practical frameworks. This systematic approach integrates empirical findings across space typologies while preserving contextual distinctions necessary for implementation. The structure of this paper is as follows. Section 2 presents a literature review examining quality factors across different public space types, organizing findings by urban spaces, open spaces, green spaces, parks and waterfronts, streets and squares, and public facilities. Section 3 outlines the methodology for developing the unified hierarchical framework, including research design, data foundation, factor analysis, conceptual clustering, distribution pattern analysis, cross-cutting factor management, and validation approaches. Section 4 presents the systematic development of the framework through seven phases, detailing the integration of multi-typology factors, cross-typology semantic analysis, distribution pattern classification, domain-informed hierarchical clustering, strategic placement of cross-cutting factors, space-type distribution analysis, and final structure generation and validation. Section 5 discusses the findings and implications. Section 6 provides the conclusion.

## 2. QUALITY FACTORS BY PUBLIC SPACE TYPOLOGY

Public space quality assessment draws from multiple theoretical frameworks that explain human-environment interactions. Lynch's theory of urban imageability establishes how spatial legibility affects navigation and understanding [12] of place. Environmental psychology, particularly Attention Restoration Theory, demonstrates how physical environments influence cognitive functioning and well-being [13]. Social theories emphasize public spaces as "third places" that foster community interaction beyond home and work [14]. Universal design principles provide frameworks for inclusive accessibility [15]. These theoretical foundations help explain why certain quality factors consistently emerge across different space types and contexts. This section examines quality factors across different public space typologies, synthesizing findings from research studies to identify patterns in assessment approaches. The analysis organizes key quality factors identified in the literature, their occurrence, and relationships across space types to inform framework development.

Before proceeding to the analysis, we establish that urban spaces are classified into two main categories: open spaces and public facilities. Open spaces are further divided into three distinct types: parks and waterfronts, streets and squares, and green spaces. Public facilities encompass a diverse range of built environments, including campuses, playgrounds, entertainment and recreation centers, historic centers, civic and community centers, municipal markets, and libraries. This classification is based on the framework established in [16]. The hierarchical classification is illustrated in Figure 1, which thus shows that urban spaces encompass all areas where people interact in cities, including outdoor spaces, shopping malls, office buildings, and residential areas. This typological division is necessary as it enables a more organized analysis and facilitates a better understanding of how quality factors vary across different space types and their specific functions.

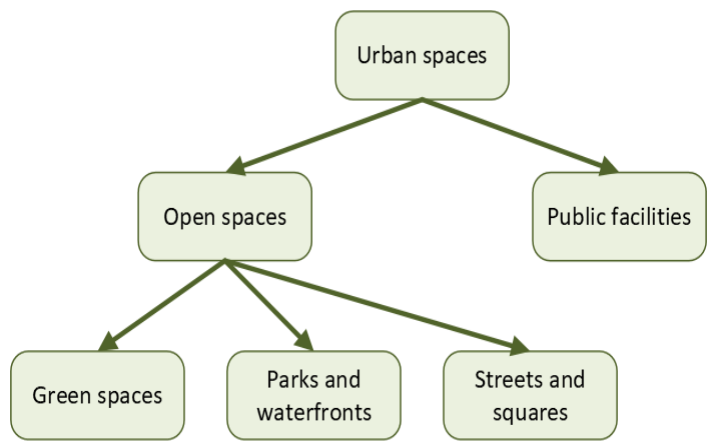

Figure 1. The hierarchy of public space categories in urban research.

## A. Urban Spaces

Urban space quality research builds on Lynch's concepts of legibility, imageability, and wayfinding, which emphasize how spatial organization affects user navigation and experience [12]. Environmental psychology suggests urban environments must balance stimulation with comprehensibility to support positive user experiences [17]. These theoretical foundations explain why accessibility, visual clarity, and spatial organization emerge as primary quality factors in urban space assessment.

Urban spaces represent the broadest category in this analysis, encompassing a diverse range of publicly accessible areas within cities. The distribution of research papers across different public space categories shows urban spaces constituting the largest portion (34%) of studies, reflecting their fundamental importance in urban research and planning. The quality factors addressed in urban space research reveal several key trends, as shown in Table 1. Accessibility features prominently across studies [18–20], highlighting the importance of physical access, connectivity, and mobility in urban environments. Comfort factors, particularly thermal, acoustic, and psychological aspects, receive considerable attention [21, 22], reflecting a growing recognition of experiential qualities in public space success. Many studies address multiple dimensions simultaneously [23, 24], indicating a trend toward more comprehensive assessment approaches. The review also reveals methodological diversity, with studies utilizing various techniques from quantitative indices [25] to qualitative assessments [26] and mixed-method approaches [27]. This methodological plurality reflects the multifaceted nature of urban space quality and the challenge of developing standardized assessment tools.

Regarding space types, the urban spaces category encompasses a wide range of environments, from general public spaces in Poland [28] to specific urban areas in Iranian cities [29]. Many studies address multiple space types collectively within urban contexts, such as research examining public spaces in Dubai [30] and urban space quality in Poznań [31]. This broad spatial coverage acknowledges the interconnected nature of public spaces within urban systems.

## B. Open Spaces

Open space quality draws from environmental psychology theories emphasizing the restorative effects of unenclosed environments [13]. Research on thermal comfort and environmental stress demonstrates how physical conditions directly influence space usage and user well-being [32]. These theoretical frameworks help explain the emphasis on comfort, environmental quality, and accessibility found across open space studies. Open spaces constitute areas characterized by the absence of buildings, providing relief from urban density and opportunities for recreation and social interaction. As seen in Table 2, open space research demonstrates a particular emphasis on environmental comfort factors, especially thermal and acoustic conditions. Studies [33] and [34] focus on the thermal aspects of open spaces. This emphasis likely stems from direct exposure to weather conditions and urban microclimates, making environmental quality a primary determinant of usability. [35] highlight the importance of soundscape quality, while [36] addressing thermal comfort evaluation methodologies. Accessibility also features prominently in open space assessment [37, 38], with particular attention to inclusive access for diverse user groups, including elderly populations [39] and during emergency situations [40]. The emphasis on inclusiveness reflects growing recognition of equity concerns in public space provision. Methodologically, open space research often employs specialized tools for specific quality dimensions, such as thermal comfort indices, acoustic measurement protocols, and accessibility evaluation frameworks. [41] proposes a comprehensive directory for multi-functional public spaces, while [42] presents a model for outdoor public space assessment that integrates multiple quality dimensions.

The typologies examined in open space research span diverse contexts, including deprived urban areas [43], town centers [38], and open public spaces in countries like Ethiopia [44]. Research approaches range from broader examinations of open space quality across cities to focused studies of specific environments like outdoor pedestrian spaces [45] and urban space in rapidly urbanizing cities [46]. This diversity of spatial contexts enables cross-cultural and cross-contextual comparison of quality factors.

## C. Green Spaces

Green space quality assessment builds on Attention Restoration Theory, which posits that natural environments reduce mental fatigue and support cognitive recovery [47]. Environmental psychology research demonstrates that contact with nature provides psychological benefits, including stress reduction and mood improvement [48]. These theoretical foundations explain why aesthetic quality, naturalness, and sensory experience emerge as central factors in green space evaluation. Green spaces, characterized by vegetation as their primary feature, serve crucial environmental and social functions within urban areas. Table 3 summarizes papers focusing on green spaces, revealing distinct quality priorities compared to other public space types. Aesthetics receives particularly strong emphasis in green space literature [49, 50], reflecting the visual importance of natural elements. Studies frequently address vegetation structure, visual quality, and landscape features as determinants of user satisfaction.

Accessibility considerations in green space research focus not only on physical access but also on equitable distribution of high-quality green areas across urban regions [51–53]. This equity dimension distinguishes green space accessibility assessment from approaches used for other public space types. The ecological quality of green spaces emerges as another distinctive focus area [54, 55], with studies examining biodiversity, environmental performance, and sustainability aspects alongside traditional social and functional qualities. This ecological emphasis reflects the unique role of green spaces in providing ecosystem services within urban environments. Research methodologies for green space assessment frequently incorporate GIS-based analyses [53], vegetation quality metrics [49], and ecological indicators [54] alongside user perception surveys [56] and observational studies [50]. This multi-method approach acknowledges both objective environmental characteristics and subjective experiential qualities.

The specific space types examined in green space research include urban vegetation landscapes [49], public urban green spaces [51], and roadside green spaces [50]. Studies also examine parks, forests, and nature reserves [53], as well as community-scale green spaces [57]. This variety highlights the different scales and configurations of green spaces within urban systems, from small-scale roadside plantings to larger park areas.

### D. Parks And Waterfronts

Parks and waterfront quality assessment integrates multiple theoretical frameworks. Recreation theory emphasizes the importance of activity diversity and programming for sustained engagement [58]. Environmental psychology research on restorative environments highlights how natural settings and water features reduce stress and support well-being [59]. Social theory emphasizes parks as community gathering spaces that build social capital [60]. These theoretical foundations explain why comfort, activity programming, and social interaction emerge as key quality factors in park evaluation. Parks and waterfronts represent specialized open spaces designed primarily for recreation, relaxation, and nature connection. Table 4 provides a summary of studies examining these space types and their quality considerations. It reveals distinct patterns in quality assessment approaches. Comfort emerges as a primary consideration [61, 62], with studies addressing both physical aspects, like thermal conditions and psychological dimensions, such as perceived safety and relaxation. The emphasis on comfort reflects the recreational purpose of these spaces, where user satisfaction depends heavily on environmental quality. Vitality also features prominently in parks and waterfront research [63, 64], focusing on usage patterns, activity diversity, and social interactions. Studies like [65, 66] employ public space indices to measure dimensions such as inclusiveness, meaningful activities, and pleasurability as indicators of vital park environments. Waterfront-specific research adds further dimensions, particularly related to the integration of water features and visual connections to aquatic environments [67–69]. These studies highlight unique quality considerations for blue spaces, including viewsheds, water quality, and edge treatments.

Methodologically, parks and waterfront research often employ standardized indices (e.g., Good Public Space Index, Public Open Space Index) alongside specialized assessment tools for environmental quality, usage patterns, and user perceptions. This methodological diversity reflects the multifunctional nature of these spaces. The specific spaces examined include urban parks [70, 71], parks for youth [72], national forest parks [73], and thematic parks [74]. Waterfront research focuses on waterfront public spaces [67], urban blue spaces [68, 69], and specific waterfront areas like the Maltepe Fill Area [13]. These diverse examples highlight the variety of park and waterfront configurations across different urban contexts.

### E. Streets And Squares

Street and square quality draws from urban design theory emphasizing the dual function of streets as movement corridors and social spaces [75]. Social life theory demonstrates how physical design influences spontaneous interaction and community vitality [76].

Safety research builds on Crime Prevention Through Environmental Design (CPTED) principles linking physical features to security perceptions [177]. These theoretical foundations explain the emphasis on walkability, safety, and social programming found in street and square studies. Streets and squares constitute the connective tissue of public space networks, serving both movement and social functions.

Table 5 summarizes research papers focusing on these space types. It demonstrates a distinctive emphasis on functionality [152–154], reflecting the primary movement role of these spaces. Studies frequently address aspects such as walkability, connectivity, and the balance between different transport modes in determining street quality. Safety emerges as another significant factor in streets and squares assessment [155], with particular attention to nighttime conditions, visibility, and security perceptions. Studies like [4] incorporate safety alongside convenience and attractiveness in global walkability assessment frameworks.

The social dimension of streets receives attention through studies of social life and vitality [156, 157], examining how physical design influences human interaction and public engagement. This social emphasis acknowledges the dual role of streets as both movement corridors and social spaces. Methodologically, streets and squares research often employs spatial syntax analysis, pedestrian counting, and observation protocols alongside user surveys and environmental measurements. This mixed-method approach reflects the complex interplay between physical infrastructure, social behaviour, and environmental conditions. The specific spaces examined range from global pedestrian streets [4] to urban streets in specific cities [158]. Studies also examine traditional commercial streets [159], streets in urban villages [46], and historic districts [160]. The diversity of street types highlights the varying functions and characteristics of these linear public spaces across different urban contexts, from pedestrianized zones to vehicle-dominated corridors.

Table 1. Summary of papers on urban spaces.

| Citation | Quality factors | Type of spaces |
|---|---|---|
| [4] | Safety, security, convenience, attractiveness, and policy support, pedestrian environments | Pedestrian spaces across cities worldwide. |
| [11] | Accessibility, multifunctionality, safety, legibility, sustainability, human scale, identity, interactivity, flexibility, scenario | Public space |
| [18] | Accessibility | Tehran, Iran |
| [19] | Pedestrian accessibility | Various public spaces within a residential radius, including local food shops, green spaces, public squares, schools, and playgrounds |
| [20] | Accessibility, spatial configuration | Public spaces |
| [21] | Comfort (temperature, humidity, mental state, busyness, location, hygro-thermograph), user preference | Public smart spaces |
| [22] | Thermal comfort, human perception, wind comfort, comfort | Publicly accessible urban spaces |
| [23] | Land occupation (housing density, urban compactness), urban complexity density and variety of activities (diversity), outdoor spaces quantification and classification of outdoor spaces features of outdoor spaces, mobility and services (modal share, public transport services, space for pedestrians and bicycles, urban services), urban metabolism energy efficiency and management, air quality and climate change), social cohesion (population social mix, affordable housing, public facilities), accessibility, diversity, concentration, contact opportunity, concentration ,diversity, need for aged buildings, distance to border vacuums, receptivity, imageability, legibility, psychological access, symbolism and memory | Urban space in Granada Spain |
| [24] | Landscape, landmarks and architecture, mobility and infrastructure, health comfort and safety, elements of public space, activities and sociability, amenities | Urban space |
| [25] | Composition/legibility/image/character/continuity and enclosure, vitality/flexibility/adaptability/use and activities/diversity, comfort/fulfillment of needs/convenience, accessibility/permeability/linkages/ease of movement, safety/control, consistency with sustainable development idea, pedestrian accessibility, bicycle accessibility, car accessibility, public transport accessibility, aesthetics, safety, cleanliness, organized attractions | Public spaces in town centers |
| [26] | Activities, attendance, communication, comfort, location | Public spaces |
| [27] | Comfort, diversity and vitality, inclusiveness, and image and likeability | Urban public spaces |
| [28] | Sensitivity, aesthetics, legibility, safety, durability, reliability, practicality, functionality | Public spaces in Poland |
| [29] | Accessibility, proximity, social activities, urban spaces, vitality | Public spaces |
| [30] | Liveable, surveillance, enclosure and privacy, human scale and density, the stage and backstage effect, surveillance, sustainable design, security | Urban space |
| [31] | Accessibility, functionality, aesthetics, safety, cleanliness | Public spaces in an urban area, Poznań |
| [77] | Financial resources/population, economical activities, environmental hygiene, subsistence, protecting the environment, access, risk/facility manage, bodily statistics, cultural statistics, cultural institutions, markets, lifestyle, technology, infrastructure, | Public spaces in Kuhdasht county Iran |
| | social commitments, variability, bank facilities, processing of resources, living wage | |
| [78] | Technological forces, social interaction, sociocultural forces, political forces, economic forces | Urban public space |
| [79] | Social issues, governance, environmental, citizen well-being, spatiality, accessibility, security, management of water resources, green spaces, sustainability, facilities, activities, biodiversity, noise exposure | Urban Public Space |
| [80] | Protection, comfort, enjoyment, quality design, comfort, liveability, quality design, aesthetics, objective characteristics of bench (dimension, proximity, materials), subjective characteristics of benches (object specific, shelter, security), aesthetics of benches (style abundance, location, design) | Urban spaces |
| [81] | Comfort | Urban public spaces |
| [82] | Green and/or landscaped areas, type of vegetation, children's, sports and animal areas, accessibility, perspective - sense of spaciousness, pedestrian circuit in the public space, design of spaces, care of spaces | Urban public spaces |
| [83] | Safety, benefit, cost, reliability, novelty, efficacy, accuracy, usability, defectiveness, looks | Pedestrian spaces in an urban area, Olsztyn |
| [84] | Safety and security, convenience and attractiveness, policy support, pedestrian environments | Pedestrian environments worldwide |
| [85] | Inclusiveness, urban scene, urban furniture, comfort, vegetation, health, streetlights, safety, walkability | Historic city center, examining streets, sidewalks, squares, parks, and pedestrian zones |
| [86] | Livability, sustainability (social sustainability, economic sustainability, environmental sustainability), walkability | Urban areas |
| [87] | Environmental pollution, activity, image and form, liveability, vitality, urbanity, mobility, sustainability | New light rail transit (LRT) |
| [88] | Characteristics of quality, safety, accessibility, mobility, urban context, user's perception of space, regeneration of public space. | Public spaces |
| [89] | Sustainable mobility, urban beauty, accessibility, liveability, quality of life, security, demographic and economic, environmental, and spatial, infrastructural and urban planning, attitudinal, environmental protection, health and safety, mobility | Urban spaces in Iranian cities |
| [90] | Perception of landscape | Public spaces |
| [91] | Accessibility, comfort, socialization, and activity | Several types of public spaces |
| [92] | Legibility, enclosure, complexity, crime potential, wildlife, and lighting, comfort, activity, environmental perception | Small public spaces |
| [93] | Thermal environment, green areas, surface reflectivity, and urban geometry | Urban areas |
| [94] | Uses and activities, accessibility and integration in the urban environment, connection to living, comfort and image, social interactions. | Public spaces |
| [95] | Urban environmental quality, air pollution, noise | Urban area |
| [96] | Urban accessibility, urban planning, design, environments, activities, attractions, maintenance, livable | Urban public spaces in northern cities |
| [97] | Physical environment, leisure environment, safety, functional variation, services, safety, shopping and work, signboard, traffic, and parking lots | Historical district |
| [98] | Accessibility, availability, affordability, participation, awareness, climate, financial safety, safety, and quality of life | City |
| [99] | Comfort, image, usage, and vitality | Urban public spaces in Chandigarh, India |
| [100] | Comfort, freedom, integration, space/user, design, activity, choice, diversity, vitality, | Public space |
| [101] | Social life and socialization, activities, access and linkage, identity, and image | Public spaces in Irkutsk, Russia |
| [102] | Social sustainability, temporal use, spatial use, social use, age diversity, gender diversity, intensity of stay, intensity of activities, activities, people, accessibility, physical characteristics, meanings, and culture values,3d visual quality, quality of the atmosphere, physical settings of activities' facilities | Public spaces |
| [103] | Control, reliability, validity, access/territoriality, laws/rules, surveillance/policing, design/image | Different urban space types |
| [104] | Publicness, identity, connectivity, spatiality, usability, environment | Urban public spaces |
| [105] | Accessibility, utility, sustainability, community, amenity, safety, constructability, functionality, integration, community, aestheticism, aesthetic visual quality | Urban public spaces in Gwangju, South Korea |
| [106] | Inclusiveness, meaningful activities, safety, comfort, and pleasurability | Public space in Ali Mendjeli, Constantine, Algeria |
| [107] | Infrastructures and mobility, commercial and industrial fabric, historic architecture, facilities, leisure and tourism, territorial green spaces, countryside, and water infrastructure | Urban public spaces in Orihuela, Spain |
| [108] | Publicness, ownership, control, civility, physical configuration, and animation | Public spaces |
| [109] | Environment, sense of place, use, protection, activities, green areas, walkable, comfort, identity, vitality, liveability, presence, and accessibility to open spaces | Urban space |
| [110] | Inclusiveness (architecture and urbanism, infrastructure, nature, health and well-being, social environment, development, and supplementary criteria) | Urban spaces |

| Citation | Quality factors | Type of spaces |
|---|---|---|
| [111] | Spatial parameters, functional distribution (activities), global integration, connectivity (of pedestrian network), visual space perception | Residential complexes |
| [112] | Urban design, liveliness, activities, accessibility, connectivity, integration | Public space |
| [113] | Space as a place of work, creates economic and functional custom, uniqueness/offer of space, property market prices, dynamics of property price fluctuation, living area/1 inhabitant, perception of business attraction, business stability, interactivity, sense of habit, relaxation, sense of belonging, intensity, continuity, bringing back memories, place identity, attractiveness (beauty of place), interesting architecture, lack of barriers, subjective assessment of place accessibility (perception of the physical and communicative accessibility of a place), management aesthetics, management functionality | Public spaces |
| [114] | Pedestrian views, ratio between resident and floating populations, diversity of uses, resident population, pedestrian comfort, permeability, spatial legibility, and continuity | Urban public spaces |
| [115] | Accessibility, management, diversity | Recreational public spaces in suburban contexts |

Table 2. Summary of papers on open spaces.

| Citation | Quality factors | Type of spaces |
|---|---|---|
| [5] | Air temperature, relative humidity, solar radiation, wind speed and direction, sun exposure and shade position, thermal comfort, temperature, greenery | Urban area |
| [33] | Thermal comfort, thermal sensation, physio-thermal environment, thermal comfort, air temperature and relative humidity, wind speed, urban environment, vegetation, urban morphology, water | Outdoor public spaces |
| [34] | Thermal comfort, urban geometry, air temperature, air speed, humidity, mean radiant temperature, metabolic rate, and clothing levels | Urban geometry in Ancona, Italy |
| [35] | Relaxation, communication, spatiality and dynamics, acoustic comfort, subjective loudness | Urban open public spaces |
| [36] | Thermal environment, comfort, quality of life, thermal comfort, temperature, solar radiation, wind speed, and relative humidity, parameters of different squares, the varying orientations of streets, and the locations of green spaces, thermal-sensation, air temperature, relative humidity, wind speed, globe temperature, thermal stress, solar radiation | Rural village public spaces |
| [37] | Inclusive, accessibility, permeability, topography, pedestrian connectivity, openness, accessible infrastructure, walkable | Pedestrian catchments |
| [38] | Composition, legibility, image, character, continuity and enclosure, vitality, flexibility, adaptability, use and activities, diversity, comfort, fulfilment of needs, convenience, accessibility, permeability, linkages, ease of movement, safety, control, consistency with sustainable development principles | Town centers |
| [39] | Accessibility, parking facilities, accessibility by urban transport systems, safety entrance and exit, physical connection with near environs, perceiving from environment, effective symbolic element for preference, necessary urban furniture, lightening, seating, trash containers, information communication panels, tents, fences, toilets, optional urban furniture, pool, plastic element, bicycle park element, kiosk, pots, clock, fountain, phone booths, design features, transparency in design, continuity in design, connection with other socio-cultural spaces, pavement, material choice, application techniques, supporting the design, maintenance services, structural quality, planting quality, management services, security, maintenance applications, organization of the socio-cultural facilities, clearness | Public open spaces |
| [40] | Inclusive, universal accessibility, pedestrian mobility, inclusive shelter, autonomy, at-risk populations, accessibility and usability, perceptions and experiences, interactions, disaster risk management | Public open spaces |
| [41] | Pleasurability, safety, comfort, desirable activities, inclusiveness, users of diverse ages, genders, physical abilities entrance controlled, diversity of activities and behaviors, differential signage, over securitization, openness and accessibility, participation of users in activities, usefulness to businesses, suitability of the spatial layout, diversity of businesses offered, availability of food, over securitization, spatial flexibility suiting user's needs, range of activities, community gathering/third places, safety from traffic volumes, safety form crimes ,day/night, over securitization, light quality, visual / physical connection or openness, appropriate maintenance & physical condition, noise pollution, elements discouraging spatial use, micro-climate (shade & shelter),street furniture & artefacts, seating areas (public / business), interestingness, attractiveness, sensory complexity (density / variety), personalization of facades, permeability of street facades, sense of enclosure, imageability | Streets, squares, parks, and in-between building spaces |
| [42] | Covering materials, environmental conditions, quality of life, social interaction enhancing, amenities and accessibility, sustainable management, visual comfort, thermal comfort, acoustic comfort, olfactory comfort, durability and suitability, adaptability, accessibility, material combination, gases, or particles emission, socialization, safety, plant species, water supply, route, maintenance, cultural identity, regional or local resources availability | Streets, squares, and parks |
| [43] | Dogs' waste, litter, broken bottles and glass, drunk people, aggressive teenagers, unsafe playground, vandalism, accessibility, playground, lack of benches, playground, walkways, grass and trees, access to other services, walking, playing on the grass, walking children, walking babies, using the playground, cycling, walking dogs, using the football pitch, sitting on the grass, playing with sand, gatherings for BBQs, skate-boarding, swinging, user perspective and use pattern, cleanliness, maintenance, safety and the open space layout, design and quality | Deprived urban areas |
| [44] | Accessibility and linkage, accessibility & linkage, use and activity, comfort and image, sociability, security, building plot, exposure, front set-backs, shopping, social services and facilities, safety & travel distance to | Rapidly urbanizing city |

|  | services, ventilation, circulation spaces, density, low built-up area ratio, density, high floor area ratio |  |
|---|---|---|
| [45] | Thermal environment, outdoor air temperature, weather, clothing, number of subjects, air temperature, humidity, and wind velocity | Urban pedestrian spaces |
| [116] | Functional (building), building arrangement, public facilities, friendly design space, signage in building, functional (access), pedestrian comfortability, regulation for disabilities, accessibility to public facilities, separated vehicle and pedestrian paths, ease of accessibility for public transport, freedom of movement in the city, visual, building facade, building variation, landscape design, human scale, street furniture, urban experience, variety activities, good pedestrian walkway, activities can be carried out until night, variety of entertainment activities, sense of place, building identity, square identity, sense of history, legibility with the surrounding area, legibility of square, urban quality, functional use, building, accessibility, visual, urban experience, sense of place, identity, structure, meaning | Public open space |
| [117] | Permeability, environmental comfort, sustainability, walkability, connectivity, access and links, active community, functional mix, sense of place, vitality, meaning, landmarks, noise, risk perception, protection | Open public spaces |
| [118] | Activities, usage, no. Of activities, amenities, children's play equipment, no. Of play equipment, picnic tables, parking facilities, public access toilets, kiosk or café, seating, clubrooms/meeting rooms, rubbish bins, dog litter bags, drinking fountains, domain: environmental quality, water features, no. Of water features, aesthetic features, no. Of aesthetic features, park size, number of trees, gardens, paths, shade along paths, watered grass, graffiti, vandalism, litter, safety, lighting, visible roads, visible houses from center, surrounded by secondary roads only | Public open spaces |
| [119] | Liveliness, accessibility, walkability | Synthetic urban environments representing intersections, streets, squares, and open spaces |
| [120] | Safety and coexistence, lighting, furniture of stay in the street, presence of pedestrian streets, presence of roads, existence of public parks or large areas of free areas near, urban elements that impede visual control, transit of people, police presence and surveillance with other security measures, spaces that favor the concentration of people, urban deterioration (lack of cleaning, deterioration of buildings and other elements of public space), main commerce use, main home use, main tourist use, diversity of uses | Urban space |
| [121] | Feeling of safety, lighting, illumination, light color temperature, uniformity, glare, city dummies, intercepts, residual deviance | Outdoor |
| [122] | Space, human perception of space, accessibility, civility, control, animation, physical configuration, ownership | A park and a neighborhood |
| [123] | Air environment, summer ventilation and shade, winter sunlight and wind blocking, barriers to use, environmental cleanliness, activity facilities, sanitation, educational facilities, commercial facilities, open air facilities, space range, motor vehicles and nonmotor vehicles, parking lots and parking spaces, convenience of transportation, night lighting, surveillance | Three relocated communities |

| | facilities, security facilities, people flow, acoustic environment, guard room and security staff, accessible parking, street aspect ratio d/h, water environment, natural landscape, artificial landscaping, community atmosphere, pleasure, security, comfort | |
|---|---|---|
| [124] | Location and proximity, accessibility, quality of open space, comfort, lighting, safety, maintenance and cleanliness, shade, and protection to enable year-round activity, image, aesthetics, facilities, activities, utilization, perceptions, opinions on priority | Three Asian cities |
| [125] | Utilization, number of parks, size of parks, accessibility, facilities and amenities, attractiveness, safety, comfortability | Public open spaces |
| [126] | Auditory comfort, visual comfort, thermal comfort, humidity, noise, glare, thermal sensation, visual sensation, auditory sensation, human perception | Public open spaces |
| [127] | Thermal-acoustic environment, comfort, relative humidity, wind speed, effective temperature | Outdoor public spaces |
| [128] | Perceptual sensation, micrometeorological and site measurements, thermal comfort, temperature, mean radiant temperature, sky view factor and aspect ratio, livability, quality of life, thermal sensation, comfort, perception, urban morphology, microclimate, relative humidity, wind speed, air temperatures, globe temperatures, environmental radiation, radiant temperature, ambient and thermal temperature, livable, shading and vegetation | Urban open spaces, parks, and streets within a city district |
| [129] | Thermal comfort, urban morphology's geometry, shape, orientation, and mass-space combination, air temperature, relative humidity, wind direction, wind speed | Urban area |
| [130] | Temperature perception, wind perception, bioclimatic comfort, age, origin, clothing, activity and motivation, air temperature, relative humidity, solar radiation, long-wave radiation, and wind speed | Public outdoor spaces |
| [131] | Soundscape, satisfaction, preference | Public spaces, including city parks, religious sites, natural scenic areas, and cultural creative industrial zones |
| [132] | Urban analysis, urban design, identity and proximity, void shape, vertical plane and permeability, environment, use and appropriation, urban system, visibility and connectivity, urban indices and density, generic labels | Public open spaces |

Table 3. Summary of papers on green spaces.

| Citation | Quality factors | Type of spaces |
|---|---|---|
| [49] | Visual aesthetic quality, vegetation landscape, visitor's perspective, vegetation structure, plant density, and height ratio | Urban vegetation landscapes |
| [50] | Quantitative public preference, visual quality, roadside green spaces, landscape features | Roadside green spaces |
| [51] | Accessibility and spatial quality, affordability, spaciousness, aesthetics | Public urban green spaces |
| [52] | Accessibility, attractiveness, facilities, quietness, culture and history, nature, space | Parks, forests, nature reserves, farmlands, and other open areas |
| [53] | Accessibility and quality | Urban green spaces - parks, forests, and nature reserves |
| [54] | Ecological, microclimatic, and social perspective, naturalness, and artificiality | Urban green spaces focusing on parks |
| [55] | Wind direction, rain management, solar orientation, micro-climate, socio cultural, activity, view and orientation, environment and location, circulation, landscape quality physical performance, dimension and function, aesthetic structure visual permeability sense of place sense of belonging tourism facility, circulation, biodiversity appearance dimension, ecosystem/ecology, structure diversity/layers of landscape, air and water qualities, environmental pollution (air and noise), diversity of vegetation and quantities, accessibility (walk and bicycle), heritage economic, interaction, social activity interaction, cultural interaction, community heritage and historical attachment, tourism | Public green open spaces |
| [56] | User perception, qualitative green space, quality of life, quietness, spaciousness, cleanliness and maintenance, facilities and feeling of safety, naturalness, historical and cultural value | Urban green space |
| [57] | Accessibility | Public green space |
| [133] | Thermal comfort, thermal humidity | Urban green open space |

Table 4. Summary of papers on parks and waterfronts.

| Citation | Quality factors | Type of spaces |
|---|---|---|
| [61] | Thermal comfort, sensation | Urban Park |
| [62] | Thermal comfort, landscape spaces, air temperature, thermal sensation, temperature, relative humidity, wind speed, global solar radiation | Urban parks |
| [63] | The intensity of use, the intensity of social activity, the duration of activity, the variations in usage, and the diversity of use | Urban Park |
| [64] | Vitality, functional mixing, travel dynamics, levels of service, road networks and population density, attractiveness | Parks and squares in medium-sized cities |
| [65] | Inclusiveness, safety, comfort, pleasurability and meaningful activities | City parks |
| [66] | Inclusiveness, meaningful activities, comfort, safety, pleasurability | District centers in Delhi, India |
| [67] | Accessibility, service capacity | Waterfront public spaces |
| [68] | Harmony, visual spaciousness & diversity, safety, sidewalk conditions, environmental management, water conditions, artificial elements, harmony, aesthetic quality, multisensory & nature, mystery | Urban blue spaces |
| [69] | Environmental aspects, transport planning, landscape architecture and management, urban design and public health, physical aspect, social aspect, aesthetic aspect | Urban blue spaces |
| [70] | Quality of life, sustainable development/prosperity, activity, social benefits, ecological benefits, economic benefits | Urban parks |
| [71] | Surrounding environment and accessibility, quietness, internal environment, water features, services and facilities, health issues, security and safety, and aesthetics | Parks in a city |
| [72] | Reliability, park size, maintenance, safety, nature, structured play diversity | Urban parks for youth |
| [73] | Tourist flow, landscape environment dimension, spatial dimension, social and humanistic dimension | National forest park |
| [74] | Inclusiveness, perception, accessibility, access to activities, user's profile, physical access, social access | Thematic parks, non-thematic parks |
| [13] | Surrounding buildings and urban integration, accessibility, activities and usage, flexibility and adaptability, amenities and enjoyment, local identity and cultural significance, aesthetic and visual appeal, functionality and purpose, sustainability, social integration, rehabilitation and urban fabric reintegration, environmental quality, open spaces and freedom of movement, collaboration and governance, historical and maritime identity, economic and social value. | Waterfront fill area |
| [134] | Individual well-being, inclusiveness, engagement, sustainable spaces, and management, physical comfort, users' visit frequency, psychological comfort, availability of elegant architecture, landscape features, attractive and pleasant views, availability of noise buffer zone, equitable access, social cohesion (social sustainability), users' freedom, engagement with space, engagement with community, economic sustainability, environmental sustainability, safety and security, cleanliness and maintenance, provision of basic facilities, users' responsibility | Parks in Nagpur, India |
| [135] | Safety, surveillance, territoriality, access control, activity support, image of the place, target hardening, socio-economical context, lighting | Urban public parks |
| [136] | Accessibility, imageability, usability, sociability | Public spaces in Bangkok |
| [137] | Sufficiency, maintenance, and growth rate, park facilities | Urban parks |
| [138] | Accessibility, views, areas, services-resources, signage, safety, lighting, cleanliness, physical order, absence of noise, presence of sensorial elements | Urban parks |
| [139] | Human well-being, spatial context, natural elements, built elements, embeddedness of green spaces, access, owned and managed, activity | Urban parks |
| [140] | Accessibility, availability, legibility, links, variety of activities, activity, sustainability, use of the space for different purposes, suitability of the space for social activities, inclusivity, interactive, suitability for recreation, safety, maintenance and cleaning, charm/attractiveness, build quality | Urban Park |
| [141] | Comfort, accessibility, activity, convenience, cleanliness, health, safety, lighting, accessibility, meanings, signage, sociability, vegetation types | Urban Park |
| [142] | Landscape elements, facilities, natural surroundings, ambiance & aesthetic, accessibility, social preferences & interaction, citizen participation & community identity, recreation & play, maintenance, space utilized, contact with nature, spaces & design, safety | Green open space assessment in Malaysia, focusing on neighborhood parks |
| [143] | Users' perception, accessibility, user service of space, facilities, emotional engagement, time conflict, activity interference, noise interference, management restriction, sense of respect, sense of safety, sense of fairness, social interaction, public participation, response to users' feedback, space democracy, cultural identity and sense of belonging, self-efficacy, economy, and social capital | Urban parks |
| [144] | Perception, visit to parks and reasons, quality of place/green spaces, health, and well-being | Urban green spaces |
| [145] | Environmental sound quality, soundscape perceptions, soundscape preferences | Urban parks |
| [146] | Soundscape evaluations, level of social interaction of users' activities, familiarity and expectations, stimulation, disruption | Public space |
| [147] | Air pollution, noise and climatic nuisances, environmental quality. | A park, an urban square, and street canyons |

| Citation | Quality factors | Type of spaces |
|---|---|---|
| [148] | Evaluation of recreational use, user preferences, sociability, entertainment, socialization, visibility, evening use, comfort and image, security, being well-kept, flora, attractiveness, accessibility, proximity, ease of access, accessible to the disabled, permeability, use-activities, number of recreational activities, food and beverage areas, playground, official ceremony | Urban public green spaces |
| [149] | Perceived safety, publicness, accessibility, permanence, liveability,, usage, appropriation | Riverfront parks |
| [150] | Intensity use, the intensity of social use, people's duration of stay, temporal diversity use, variety of use, and diversity of users. | Waterfront public open space |
| [151] | Historical-cultural heritage richness and identity, functional diversity, accessibility, vitality, and spatial quality | Urban square |

Table 5. Summary of papers on streets and squares.

| Citation | Quality factors | Type of spaces |
|---|---|---|
| [4] | Safety and security, convenience and attractiveness, and policy support | Global streets |
| [5] | Safety, convenience, legibility, comfort, inspiration, and liveability | Urban streets |
| [46] | Accessibility, protected property, comfort, and enjoyment | Urban village |
| [152] | Density, mix, and access | Walkable urban environments |
| [153] | Cleanliness, comfort, and traffic | Street space |
| [154] | Street space carrying capacity, street space vitality, street environment comfort, street travel safety, and crowd social interaction | Street space |
| [155] | Safety and security, visibility and recognition, pedestrian usage, and environmental perception | Street |
| [156] | Access and linkage, comfort, and image, uses and activities, and sociability | Street |
| [157] | Physical dimension, activity dimension, social dimension, and meaning dimension | Responsive public spaces |
| [158] | Continuity, diversity, esthetic, comfort, and scale | three cities, focusing on city center streets |
| [159] | Street quality and neighborhood communication | Traditional commercial street in Belgium |
| [160] | Functional affordances, accessibility, emotional affordances, and social affordances | Historic district |
| [161] | Design, maintenance, activities, accessibility, and safety | Pedestrian spaces |
| [162] | Accessibility, safety, comfort, aesthetics, usage diversity, user diversity, and topography | Pedestrianized urban streets |
| [163] | Infrastructure quality, attraction, and structure | Street vending activity in urban areas |
| [164] | Physical visual quality, perceived visual quality, and temporal variation | Street spaces |
| [165] | Street inclusiveness and safety | Street |
| [166] | Frontage permeability, density, plot size, and network centrality | urban environments |

Table 6. Summary of papers on public facilities.

| Citation | Quality factors | Type of spaces |
|---|---|---|
| [6] | Spatial accessibility, spatial equity, playground quality, demographic need, and social need | Public playgrounds |

| [167] | Environment, mobility and parking, safety, urban space, and support services | University campuses |
| --- | --- | --- |
| [168] | Indoor environment, safety, and maintenance; furniture, utilities, and spaces; and privacy, appearance, and surrounding areas | Campus facilities |
| [169] | Indoor air quality, thermal comfort, visual comfort, acoustic quality, and electromagnetic pollution | Office buildings |
| [170] | Visual dimension, perceptual dimension, and social dimension | Campus pathway |
| [171] | Physical landscape dimension, social dimension, and emotional and spiritual dimension | Commercial public spaces |
| [172] | Physical characteristics and the socioeconomic status | Public recreational spaces |
| [173] | Visual & morphological, functional, social, perceptual & experiential, and ecological | Public courtyard quality |
| [174] | Historical value, aesthetic value, and functional value | Historical public spaces |
| [175] | Democratic, meaningful, and responsive, intensity of use, intensity of social use, people's duration of stay, temporal diversity of use, variety of use, and variety of users | Non-green public spaces in a fisher settlement, focusing on a playground |
| [176] | Accessibility | Built environment |

## F. Public Facilities

Public facility quality assessment builds on environmental psychology research demonstrating how indoor environmental quality affects cognitive performance and user satisfaction [178]. Universal design theory emphasizes equitable access to services regardless of physical ability [179]. Organizational psychology research shows how physical environments influence service delivery and user experience [180]. These theoretical foundations explain why indoor environmental quality, accessibility, and functional design dominate public facility assessment. Public facilities include libraries, community centers, and other built environments that provide specialized services and activities. Table 6 summarizes research papers examining the quality of these spaces. It shows a distinctive quality profile compared to outdoor public spaces. Functionality receives particularly strong emphasis [167, 168], reflecting the service-oriented nature of these environments. Studies frequently address aspects such as interior layout, equipment, utilities, and operational efficiency as primary determinants of facility quality. Indoor environmental quality emerges as a specialized focus area [169] with studies examining air quality, thermal comfort, visual comfort, and acoustic conditions. This indoor focus distinguishes public facilities assessment from approaches used for outdoor space types.

The social dimension of public facilities receives attention through studies of user interaction and community engagement [170, 171], examining how spatial configuration and programmatic elements influence social behaviour and institutional identity. This social emphasis acknowledges the role of public facilities as community anchors and service providers. Methodologically, public facilities research often employs post-occupancy evaluation, user satisfaction surveys, and environmental monitoring alongside spatial analysis and observation protocols. This diverse methodological approach reflects the complex interplay between building performance, service delivery, and user experience. The space types examined include university campuses [167], campus facilities [168], campus pathways [170], public recreational spaces [172], playgrounds [6], public courtyards [173], office buildings [169] commercial public spaces [171], and historical public spaces [174]. This diversity reflects the broad range of built facilities that serve public functions across urban contexts.

## 3. METHODOLOGY

This section presents the methodology used in this research to construct a comprehensive unified hierarchical quality factors framework. It presents the flow of algorithms and its inputs and outputs at each phase. The methodology flowchart is presented in Figure 2.

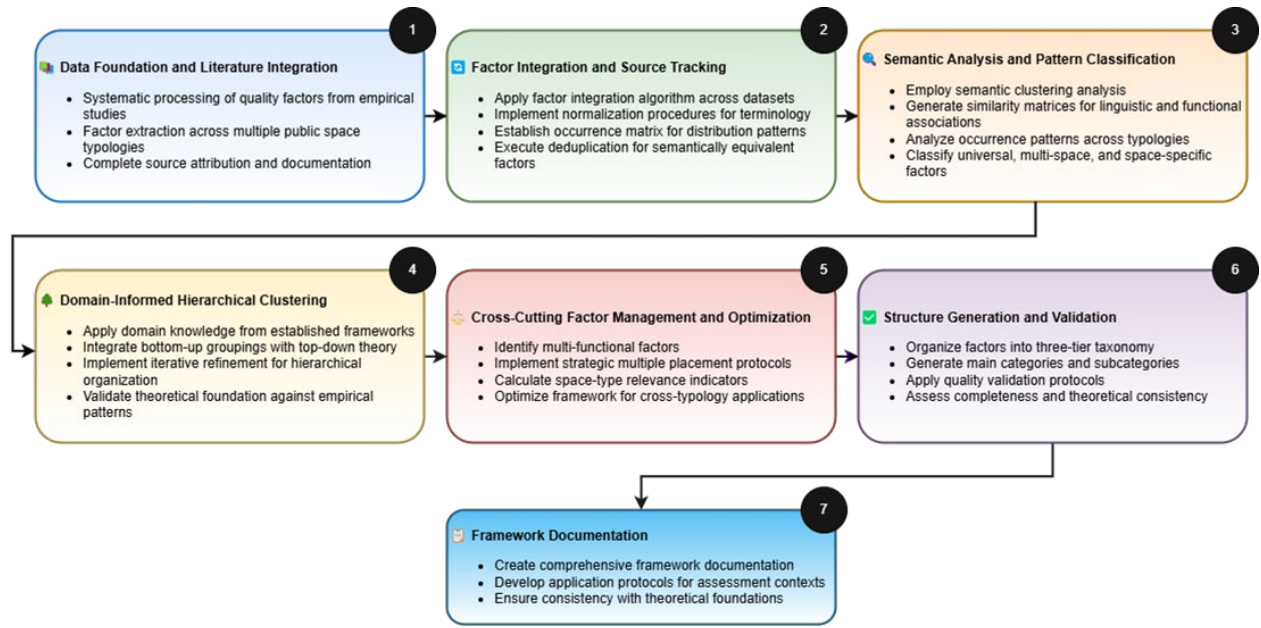

Figure 2. Methodology flowchart, listing the different stages of the data-driven domain-informed hierarchical public space quality assessment framework.

### A. Research Design and Approach

This study employed a multi-phase taxonomic development methodology to create a unified hierarchical framework for public space quality assessment. The research integrated systematic literature synthesis, multi-dataset factor integration, semantic analysis, and domain-informed clustering to transform disparate quality factors from six distinct public space typologies into a coherent three-tier taxonomic structure. The 7-phase methodology addresses complexities in public space research that simpler approaches may not adequately resolve. Phase 1 addresses terminological variations where identical concepts appear with different names across studies (e.g., 'accessibility,' 'access,' 'barrier-free'). Phase 2 examines semantic relationships that basic keyword matching often misses. Phase 3 classifies distribution patterns to distinguish universal from space-specific factors. Phase 4 incorporates domain knowledge integration, as purely computational clustering tends to produce theoretically inconsistent groupings. Phase 5 addresses cross-cutting factors that serve multiple functions simultaneously. Phase 6 examines space-type applicability based on empirical occurrence patterns. Phase 7 validates the framework structure. Each phase appears to address a methodological challenge that could compromise framework validity if omitted or simplified. The methodology followed established taxonomic development principles while incorporating cross-typology integration algorithms designed for the public space quality domain to address complexities that simpler approaches may not adequately resolve.

### B. Data Foundation and Systematic Literature Integration

The methodological foundation was established through a comprehensive domain synthesis approach that systematically processed empirical quality factors extracted from peer-reviewed studies across six major public space typologies: parks & waterfronts, streets & squares, urban spaces, green spaces, open spaces, and public facilities. The literature corpus was assembled through systematic selection processes that identified 157 empirical studies focusing specifically on public space quality assessment methodologies and factor identification. This corpus builds upon our previous systematic review [9], but applies novel algorithmic approaches to transform the identified quality factors into structured taxonomic frameworks.

The systematic extraction process involved factor identification protocols that captured quality assessment variables, indicators, and criteria employed across the selected studies. Each extracted factor was documented with source attribution and typological context, creating six datasets representing the quality assessment landscapes of each public space category. This approach appears to ensure coverage of quality factors while maintaining transparency regarding factor origins and disciplinary contexts.

### C. Multi-Dataset Integration and Source Tracking Protocol

The integration of quality factors across the six typologies employed a sophisticated Multi-Typology Factor Integration Algorithm (see section IV. A) designed to systematically combine disparate factor vocabularies while preserving complete source tracking information. This algorithm processed each typological dataset sequentially, implementing advanced normalization procedures to address terminological variations while maintaining semantic integrity.

The source tracking protocol established a comprehensive occurrence matrix that documented the precise distribution patterns of each quality factor across all six space typologies. This tracking system generated detailed source codes indicating factor frequency and typological occurrence patterns, enabling subsequent analysis of universal versus space-specific quality characteristics. The integration process implemented intelligent deduplication procedures that identified semantically equivalent factors appearing with slight terminological variations, consolidating these while preserving their complete occurrence histories across all space types.

## D. Cross-Typology Semantic Analysis and Pattern Recognition

Following successful integration, the methodology employed Cross-Typology Semantic Clustering Analysis (detailed in section IV.B) to explore potential conceptual relationships among the normalized quality factors. This analysis attempted to combine computational semantic similarity assessment with empirical co-occurrence pattern analysis to generate similarity matrices that could capture both linguistic relationships and functional associations across different public space contexts. The semantic analysis incorporated multiple similarity calculation approaches, including distributional semantic analysis based on factor co-occurrence patterns and contextual similarity assessment derived from space-type distribution characteristics. Factors that demonstrated similar occurrence patterns across space typologies were considered potentially related, while those exhibiting complementary distribution patterns were examined for possible cross-cutting functional relationships. The computational approach sought to understand how quality factors might cluster together based on both semantic meaning and empirical distribution patterns. This dual methodology aimed to identify factors that appeared to co-occur consistently across different public space types, as well as those that seemed to demonstrate complementary functional relationships. The resulting similarity matrices provided a preliminary foundation for understanding both direct conceptual connections and broader thematic relationships among quality factors across diverse urban contexts.

A critical component of the framework development involved Distribution Pattern Classification Analysis (see section IV. C) to systematically categorize quality factors based on their universality and space-type specificity. This classification algorithm analyzed factor occurrence patterns across the six typologies to identify universal factors (appearing across five or more space types), multi-space factors (appearing in three to four typologies), and space-specific factors (unique to particular typological contexts).

The pattern classification process incorporated distribution entropy analysis to assess the evenness of factor occurrence across space types, identifying factors with uniform distribution patterns as universal quality concerns and those with concentrated distributions as specialized typological requirements. Additionally, the analysis identified cross-cutting factors that serve multiple functional purposes across different quality domains, requiring strategic placement within the hierarchical framework.

## E. Domain-Informed Hierarchical Clustering

The core taxonomic development employed Multi-Level Domain-Informed Clustering (detailed in section IV.D) that sought to systematically organize quality factors into hierarchical structures based on both empirical clustering patterns and established theoretical frameworks from public space research. This clustering approach attempted to integrate bottom-up analysis of natural factor groupings with top-down application of domain knowledge derived from environmental psychology, urban design theory, and accessibility research.

The domain knowledge integration involved a comprehensive synthesis of established theoretical frameworks including environmental comfort research, safety and security theories, accessibility principles, and social interaction models. This theoretical foundation was systematically cross-referenced with empirical factor patterns to help ensure that the resulting hierarchical structure could reflect both scholarly consensus and evidence-based relationships observed in the integrated datasets. The clustering algorithm implemented iterative refinement procedures that progressively organized factors into what appeared to be main categories, subcategories, and individual factor levels.

Each level of the hierarchy was examined through semantic coherence analysis and theoretical consistency assessment to evaluate logical organization and practical utility for quality assessment applications.

### F. Cross-Cutting Factor Strategic Placement

The methodology incorporated Strategic Cross-Cutting Factor Management (see section IV. E) to address quality factors that serve multiple functional purposes and appropriately belong in several categories within the hierarchical framework. This algorithm systematically identified factors with multi-functional characteristics and implemented strategic multiple placement protocols where semantically justified. The cross-cutting factor analysis employed functional diversity assessment to identify factors serving multiple purposes simultaneously, such as lighting serving safety, comfort, and infrastructure functions. Rather than forcing such factors into single categories, the framework implementation allows strategic multiple placements with clear primary and secondary designation protocols, maintaining categorical organization while acknowledging the interconnected nature of public space quality factors.

### G. Space-Type Distribution Optimization

The framework development attempted to incorporate Space-Type Distribution Analysis (see section IV. F) to optimize the hierarchical structure based on factor occurrence patterns across different public space typologies. This analysis sought to systematically calculate space-type relevance indicators for each factor and subcategory, generating distribution profiles that could inform practical application protocols. The optimization process attempted to implement adaptive categorization procedures that adjust factor placement based on space-type distribution characteristics. Factors demonstrating universal distribution patterns were tentatively designated with universal applicability indicators, while those showing concentrated occurrence patterns appeared to receive space-specific relevance designations. This optimization aimed to ensure that the framework could potentially support both comprehensive cross-typology assessment and targeted space-specific evaluation protocols.

### H. Hierarchical Structure Generation and Comprehensive Validation

The final framework synthesis employed Comprehensive Structure Generation algorithms (see Section IV. G) that systematically organized the optimized factor clusters into a complete three-tier hierarchical taxonomy. This process generated main categories representing fundamental dimensions of public space quality, subcategories organizing related factors into coherent functional groupings, and individual factor levels maintaining complete source tracking and space-type relevance information. The structure generation process sought to implement comprehensive metadata integration that aimed to preserve complete factor provenance, occurrence frequency, and typological distribution characteristics. The completed hierarchical framework underwent Comprehensive Quality Validation (see Section IV. G) to evaluate logical consistency, conceptual clarity, and complete factor accountability. This validation process employed multiple assessment protocols including completeness verification, semantic coherence analysis, and theoretical consistency evaluation. The validation process incorporated theoretical consistency assessment that evaluated the alignment between the empirically derived hierarchical structure and established theoretical frameworks from public space research, working to confirm that the framework could reflect scholarly consensus while potentially extending theoretical understanding through comprehensive empirical integration.

### I. *Framework Documentation and Application Protocols*

The final phase involved comprehensive documentation of the unified hierarchical framework, including detailed specification of all categories, subcategories, and individual factors with complete source tracking information and space-type applicability guidelines. The framework documentation incorporates application protocols that support multiple assessment contexts, from comprehensive public space evaluation to targeted typology-specific analysis. The documented framework provides detailed guidance for factor selection based on assessment objectives, space-type characteristics, and evaluation scope requirements. This documentation ensures that researchers and practitioners can effectively apply the hierarchical framework while maintaining consistency with the underlying theoretical and empirical foundations that inform its structure and organization.

## 4. Systematic Development of the Unified Hierarchical Quality Factors Framework

The systematic development of the unified hierarchical framework follows a 7-phase algorithmic methodology that transforms quality factors extracted from six public space typologies (see the Excel files in the supplementary files) into a coherent taxonomic structure. Figure 3 illustrates the complete implementation process, showing how quality factor datasets, domain knowledge, and literature co-occurrence data integrate through sequential algorithms to produce the final validated three-tier hierarchical framework. The implementation process applies seven distinct algorithms across seven phases, each addressing specific aspects of factor integration, semantic analysis, classification, clustering, cross-cutting placement, optimization, and validation. The detailed implementation of each phase is presented in the sections below.

## A. Phase 1: Multi-Typology Factor Integration and Source Tracking Implementation

In public space research, researchers describe the same quality concepts using different terms across multiple typologies, creating integration challenges. The Multi-typology Factor Integration Algorithm (Algorithm 1) handles this vocabulary variation while preserving meaningful distinctions between factors. For example, the concept of "physical accessibility" appeared as "accessibility", "Accessibility", "access", "Access", and "physical access" across the various spatial types. These variations extended beyond capitalization to include differences in meaning and context. Similarly, while "safety" and "Safety" referred to the same concept, terms like "street travel safety" were given separate categorization due to their focused usage in transportation-oriented studies. The integration algorithm sought to process each dataset by identifying raw factor names, applying normalization rules, and checking for matches after obtaining the normalized factors. The normalized factor was then added to the integrated list if it was not already present, working to create an integrated factor list that would contain unique factors.

Spatial source tracking with notation [P×n, S×n, U×n, G×n, O×n, F×n] was also generated, which aimed to allow for understanding the empirical spread of each factor (how widely distributed each factor appeared across the different spatial typologies based on actual observed data). This approach sought to enable better interpretation in the framework development process. Moreover, source tracking worked to identify patterns across the dataset, attempting to distinguish factors that appeared to occur universally (in five or more typologies), those appearing across multiple space types (three to four), and those that seemed to remain typology-specific (limited to one or two).

Table 7 presents the inputs to Algorithm1, and its sequential outputs at different stages (inputs are shown in blue colour column, intermediate results as orange and outputs as green).

Safety appeared across five of six datasets, suggesting near-universal concern absent only from Green Spaces research. For Accessibility, the algorithm worked to normalize variations from "access" to canonical "accessibility" form. Multiple mentions (×4, ×2) in Open Spaces and Public Facilities suggested heightened emphasis in facility-focused research. Street travel safety demonstrated the algorithm's decision-making capability - despite semantic similarity to safety, its transportation-specific context appeared to require separate preservation, reflecting attempts to balance consolidation with meaning preservation. Comfort emerged as a shared factor, while thermal and physical comfort seemed to retain separate status. Lighting consolidated from multiple sources, while water features appeared to remain park specific.

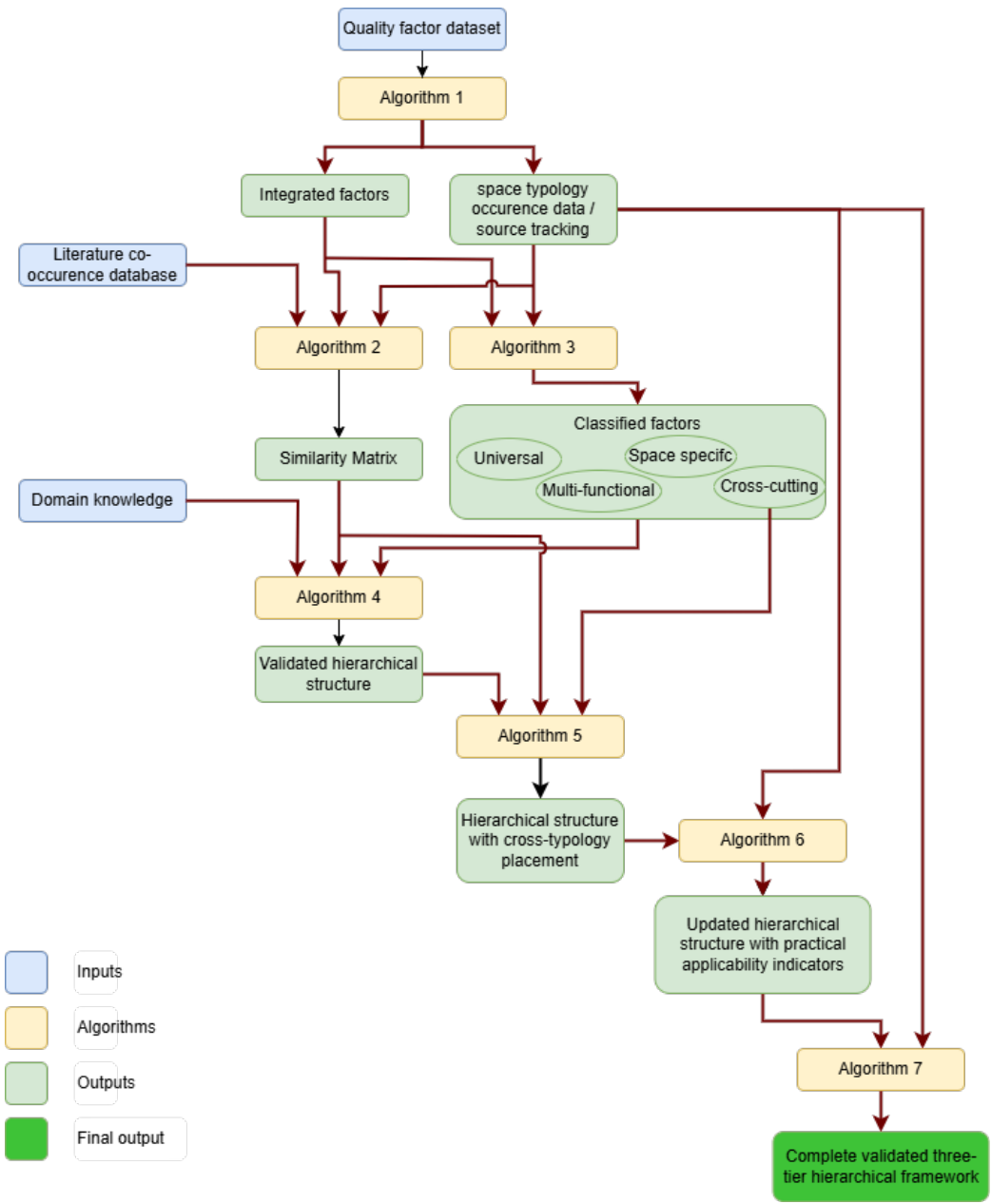

Figure 3. Framework development process.

Overall, the algorithm goes through all the 1,207 factors (steps), to reduce them into 1,029 unique factors, achieving a 14.7% reduction in redundancy. This semantic deduplication maintains meaning while simplifying the factor set for analysis. In the next phase the relationships between these unique factors are determined.

**ALGORITHM 1: Multi Typology Factor Integration**

**Input**: Six datasets with original quality factors, normalization rules

**Output**: Integrated factor list with comprehensive source tracking

1. **initialize**:
   - integrated_factors = []
   - source_tracking = {}
   - space_type_codes = ['P', 'S', 'U', 'G', 'O', 'F']
   - normalization_rules = load_semantic_equivalency_rules()
2. **for** each dataset_file, space_code **in** zip(datasets, space_type_codes):
   a. factors_list = extract_factors_from_excel(dataset_file)
   b. **for** each raw_factor **in** factors_list:
      i. normalized_factor = apply_normalization_rules(raw_factor)
      ii. IF normalized_factor NOT in source_tracking:
         - source_tracking[normalized_factor] = initialize_tracking_vector()
         - integrated_factors.append(normalized_factor)
      iii. source_tracking[normalized_factor][space_code] += 1
3. **return** integrated_factors, source_tracking

Table 7. Step-By-Step Processing Example Shows Algorithm 1's Integration of Quality Factors with Source Tracking.

| Step | Inputs | | Intermediate Results | Outputs | |
|---|---|---|---|---|---|
| | Dataset | Raw Factor | Normalized Factor | Integrated Factors List | Source Tracking Update |
| 1 | P | safety | safety | [safety] | safety: {P:1, S:0, U:0, G:0, O:0, F:0} |
| 2 | S | safety | safety (match) | [safety] | safety: {P:1, S:1, U:0, G:0, O:0, F:0} |
| 3 | U | safety | safety (match) | [safety] | safety: {P:1, S:1, U:1, G:0, O:0, F:0} |
| 4 | G | — | — | [safety] | safety: {P:1, S:1, U:1, G:0, O:0, F:0} |
| 5 | O | safety | safety (match) | [safety] | safety: {P:1, S:1, U:1, G:0, O:1, F:0} |
| 6 | F | safety | safety (match) | [safety] | safety: {P:1, S:1, U:1, G:0, O:1, F:1} |
| 7 | P | access | accessibility | [safety, accessibility] | accessibility: {P:1, S:0, U:0, G:0, O:0, F:0} |
| 8 | S | accessibility | accessibility (match) | [safety, accessibility] | accessibility: {P:1, S:1, U:0, G:0, O:0, F:0} |
| 9 | U | Accessibility | accessibility (normalized) | [safety, accessibility] | accessibility: {P:1, S:1, U:1, G:0, O:0, F:0} |
| 10 | G | — | — | [safety, accessibility] | accessibility: {P:1, S:1, U:1, G:0, O:0, F:0} |
| 11 | O | accessibility ×4 | accessibility (match) | [safety, accessibility] | accessibility: {P:1, S:1, U:1, G:0, O:4, F:0} |
| 12 | F | Accessibility ×2 | accessibility (normalized) | [safety, accessibility] | accessibility: {P:1, S:1, U:1, G:0, O:4, F:2} |
| 13 | S | street travel safety* | street travel safety | [safety, accessibility, street travel safety] | street travel safety: {P:0, S:1, U:0, G:0, O:0, F:0} |
| 14 | U | comfort | comfort | [safety, accessibility, street travel safety, comfort] | comfort: {P:1, S:0, U:1, G:1, O:1, F:1} |
| 15 | O | thermal comfort ×2 | thermal comfort | [safety, accessibility, street travel safety, comfort, thermal comfort] | thermal comfort: {P:1, S:0, U:0, G:0, O:1, F:1} |
| 16 | P | physical comfort | physical comfort | [safety, accessibility, street travel safety, comfort, thermal comfort, physical comfort] | physical comfort: {P:1, S:0, U:0, G:0, O:0, F:0} |
| 17 | S | lighting | lighting | [safety, accessibility, street travel safety, comfort, thermal comfort, physical comfort, lighting] | lighting: {P:1, S:1, U:0, G:0, O:1, F:0} |
| 18 | P | water features | water features | [safety, accessibility, street travel safety, comfort, thermal comfort, physical comfort, lighting, water features] | water features: {P:1, S:0, U:0, G:0, O:0, F:0} |
| 19 | P | security | security | [safety, accessibility, street travel safety, comfort, thermal comfort, physical comfort, lighting, water features, security] | security: {P:1, S:0, U:1, G:0, O:0, F:0} |
| 20 | S | visibility | visibility | [safety, accessibility, street travel safety, comfort, thermal comfort, physical comfort, lighting, water features, security, visibility] | visibility: {P:1, S:1, U:0, G:0, O:1, F:0} |
| 21 | G | biodiversity | biodiversity | [safety, accessibility, street travel safety, comfort, thermal comfort, physical comfort, lighting, water features, security, visibility, biodiversity] | biodiversity: {P:1, S:0, U:0, G:1, O:0, F:0} |

*Note: street travel safety is context-specific and semantically related to safety but preserved as a distinct factor due to its limited distribution and transportation-focused relevance.

## B. Phase 2: Cross-Typology Semantic Analysis And Enhanced Similarity Assessment Implementation

With the integrated factors established from Phase 1, the next step toward creating a unified framework requires clustering these factors. This step analyzes the similarity between factors and constructs a similarity matrix. Traditional methods consider only the linguistic similarity dimension and do not account for other dimensions such as functional similarity and co-occurrence patterns in the literature, preventing identification of the full relationship pattern.

The Cross-Typology Semantic Analysis Algorithm (Algorithm 2) processes all possible factor pairs from the 1,029 integrated factors, generating a 1,029×1,029 similarity matrix that captures multiple dimensions of factor relationships: linguistic similarity, distributional similarity, and co-occurrence strength. Semantic similarity combines three measures: lexical similarity (shared terminology), distributional similarity (co-occurrence patterns across space types), and contextual similarity (functional relationships within studies). The composite score weights these components equally to capture both linguistic and functional relationships between quality factors (detailed algorithms are provided in Appendix 3).

The linguistic similarity component uses computational semantic analysis to identify factors with related meanings. The distributional similarity component leverages the cross-typology occurrence patterns documented during integration. The co-occurrence strength assessment draws upon literature analysis to identify factors that appear together in empirical studies. This approach captures the range of factor relationships present in the integrated dataset. Table 8 presents Algorithm 2's operation on combinations of integrated factors, showing how different similarity dimensions contribute to relationship scores. For related factors like "safety" and "security," linguistic analysis recognizes shared semantic fields, distributional analysis identifies overlapping space-type patterns, and co-occurrence analysis confirms frequent joint appearance in literature. These dimensions combine to produce high similarity scores. For unrelated factors like "traffic" and "biodiversity," the algorithm demonstrates minimal relationships across all dimensions, resulting in low similarity scores. For functionally related factors like "lighting" and "visibility," the algorithm reveals cause-effect relationships despite different semantic domains. Distributional alignment indicates shared functional contexts, while moderate co-occurrence in lighting design literature confirms practical relationships. This approach identifies both semantic and functional relationships that single-dimension methods miss. The similarity matrix appears in Table 8, showing the first 9 entries of the integrated factors due to space limitations. The analysis identifies 529,506 unique factor pairs (1,029 × 1,029 - 1,029 diagonal elements / 2), each receiving a similarity score between 0 and 1. The diagonal is omitted as it represents the factor repeated with itself.

A threshold sought to cluster the scores into three categories. Scores above 0.75 were classified as high similarity factor clusters, scores between 0.5 and 0.75 fell into moderate similarity functional associations, and scores below 0.5 were categorized as low similarity distinct domains. The 0.75 threshold aimed to represent a conservative approach, ensuring high semantic similarity while preventing over-clustering. Values below 0.70 appeared to produce excessive false positives, grouping semantically distinct factors (e.g., 'thermal comfort' with 'social comfort'). Values above 0.80 seemed to create under-clustering, failing to group clear semantic equivalents (e.g., 'accessibility' and 'barrier-free access'). The 0.75 threshold appeared to achieve optimal balance, validated through iterative testing on factor subsets.

Table 8. Multi-dimensional similarity analysis examples showing Algorithm 2's relationship assessment across three analytical dimensions.

| Factor Pair | Linguistic Similarity | Distributional Similarity | Co-occurrence Strength | Final Score | Interpretation |
|---|---|---|---|---|---|
| Safety; security | 0.85 (shared protection domain) | 0.53 (partial overlap: P, U) | 0.91 (89% co-occurrence) | 0.77 | Strong semantic relationship |
| Lighting; visibility | 0.62 (cause-effect domains) | 0.88 (perfect alignment: P, S, O) | 0.74 (67% co-occurrence) | 0.72 | Strong functional relationship |
| Traffic; biodiversity | 0.08 (distinct domains) | 0.05 (no overlap: S, U vs G, P) | 0.02 (<5% co-occurrence) | 0.06 | Minimal relationship |
| Accessibility; access | 0.94 (direct terminology) | 0.85 (similar patterns) | 0.88 (frequent pairing) | 0.91 | Terminological equivalence |
| Thermal comfort; temperature | 0.95 (measurement relation) | 0.92 (identical: P, O) | 0.89 (direct correlation) | 0.93 | Direct measurement relationship |

The analysis identified 2,847 factor pairs (0.54% of total pairs) with high similarity scores, representing semantic and functional relationships. The Accessibility Cluster appeared to demonstrate coherence, with terminological relationships achieving scores above 0.85. The Safety and Security Cluster exhibited high scores, suggesting protection-focused relationships and appearing to confirm expectations about crime prevention and safety concepts.

The Comfort Domain Cluster exhibits user experience relationships, with measurement-related factors showing high similarity scores. The analysis identifies 15,234 factor pairs (2.88% of total pairs) with moderate similarity scores, representing functional relationships. These include lighting and visual environment relationships, natural elements functional groups, and activity programming relationships. These moderate relationships often reflect cause-effect or implementation relationships rather than direct semantic equivalence. Most factor pairs (96.58%) demonstrate low similarity scores, confirming the presence of distinct conceptual domains within the integrated dataset. Cross-domain comparisons reveal minimal relationships, while space-type specific factors show isolation patterns consistent with their specialized nature. The similarity score distribution appears in the heatmap shown in Figure 4. The algorithm thus processed all 529,506 factor pairs to generate a relationship matrix which could now assist in domain-informed clustering in Phase 4.

```
ALGORITHM 2: Cross Typology Semantic Analysis
Input: integrated factors list, source tracking data, literature co-occurrence database
Output: enhanced similarity matrix with multi-dimensional factor relationships
1. initialize:
   - similarity_matrix = create_matrix(len(integrated_factors), len(integrated_factors))
   - semantic_weights = {linguistic: 0.5, distributional: 0.3, co_occurrence: 0.2}
   - similarity_threshold = 0.75
2. for each factor_pair (factor_i, factor_j) in all_factor_combinations:
   a. linguistic_similarity = calculate_semantic_similarity(factor_i, factor_j)
   b. distributional_similarity = assess_space_distribution_patterns(
        source_tracking[factor_i], source_tracking[factor_j])
   c. co_occurrence_strength = evaluate_literature_co_occurrence(factor_i, factor_j)
   d. enhanced_similarity = weighted_integration(
        linguistic_similarity * semantic_weights.linguistic,
        distributional_similarity * semantic_weights.distributional,
        co_occurrence_strength * semantic_weights.co_occurrence)
   e. similarity_matrix[i][j] = enhanced_similarity
3. identify high_similarity_clusters where similarity > similarity_threshold
4. validate similarity_patterns against theoretical_frameworks
5. return validated_similarity_matrix
```

## C. Phase 3: Distribution Pattern Classification And Universal Factor Identification Implementation

With the similarity matrix established from Phase 2, Phase 3 sought to address the question of factor universality versus space-type specificity within public space quality assessment. This classification process aimed to determine which quality factors appeared to represent universal human needs that transcend specific space typologies versus those that seemed to reflect specialized requirements of particular public space contexts. The Distribution Pattern Classification Algorithm (Algorithm 3) worked to process each of the 1,029 integrated factors through systematic distribution analysis, examining their occurrence patterns across the six space typologies. The algorithm aimed to incorporate both quantitative distribution metrics (occurrence count, distribution entropy) and qualitative functional analysis (cross-cutting domain identification) to generate factor classifications that sought to help in subsequent clustering procedures. The analysis helps to identify universal, multi-space, space-specific and cross-cutting factors from the list of integrated factors, and its occurrence data. Table 9 presents Algorithm 3's operation on the integrated factors, showing how distribution patterns and functional analysis contribute to classifications.

For universal factors like "safety," the analysis reveals presence across five of six space types with high distribution entropy, indicating even distribution across active space types. The functional domain identification process recognizes safety as belonging primarily to the Safety and Security domain. The algorithm classifies "safety" as UNIVERSAL based on its appearance in 5+ space types. Safety represents a security concern applicable across all public space contexts, with implementation variations such as lighting strategies for parks versus traffic management for streets.

For multi-space factors like "lighting," the occurrence pattern analysis reveals presence in three space types: (P, S, O). The absence from (U, G, F) reflects the factor's concentration in outdoor public spaces requiring nighttime illumination. The functional domain identification process reveals lighting's multi-dimensional nature, with the primary domain as Comfort and secondary domains including Safety and Infrastructure. The algorithm classifies "lighting" as MULTI-SPACE with CROSS-CUTTING status due to its multiple functional roles.

For space-specific factors like "water features," the occurrence pattern analysis reveals exclusive presence in (P), representing the most concentrated distribution pattern possible. The functional domain identification process recognizes water features as belonging primarily to the Natural Elements domain. The algorithm classifies "water features" as SPACE-SPECIFIC based on its appearance in only one space type, reflecting specific relevance to parks and waterfront environments.

Table 9. Multi-dimensional similarity analysis examples showing Algorithm 2's relationship assessment across three analytical dimensions.

| Factor Pair | Linguistic Similarity | Distributional Similarity | Co-occurrence Strength | Final Score | Interpretation |
|---|---|---|---|---|---|
| Safety; security | 0.85 (shared protection domain) | 0.53 (partial overlap: P, U) | 0.91 (89% co-occurrence) | 0.77 | Strong semantic relationship |
| Lighting; visibility | 0.62 (cause-effect domains) | 0.88 (perfect alignment: P, S, O) | 0.74 (67% co-occurrence) | 0.72 | Strong functional relationship |
| Traffic; biodiversity | 0.08 (distinct domains) | 0.05 (no overlap: S, U vs G, P) | 0.02 (<5% co-occurrence) | 0.06 | Minimal relationship |
| Accessibility; access | 0.94 (direct terminology) | 0.85 (similar patterns) | 0.88 (frequent pairing) | 0.91 | Terminological equivalence |
| Thermal comfort; temperature | 0.95 (measurement relation) | 0.92 (identical: P, O) | 0.89 (direct correlation) | 0.93 | Direct measurement relationship |

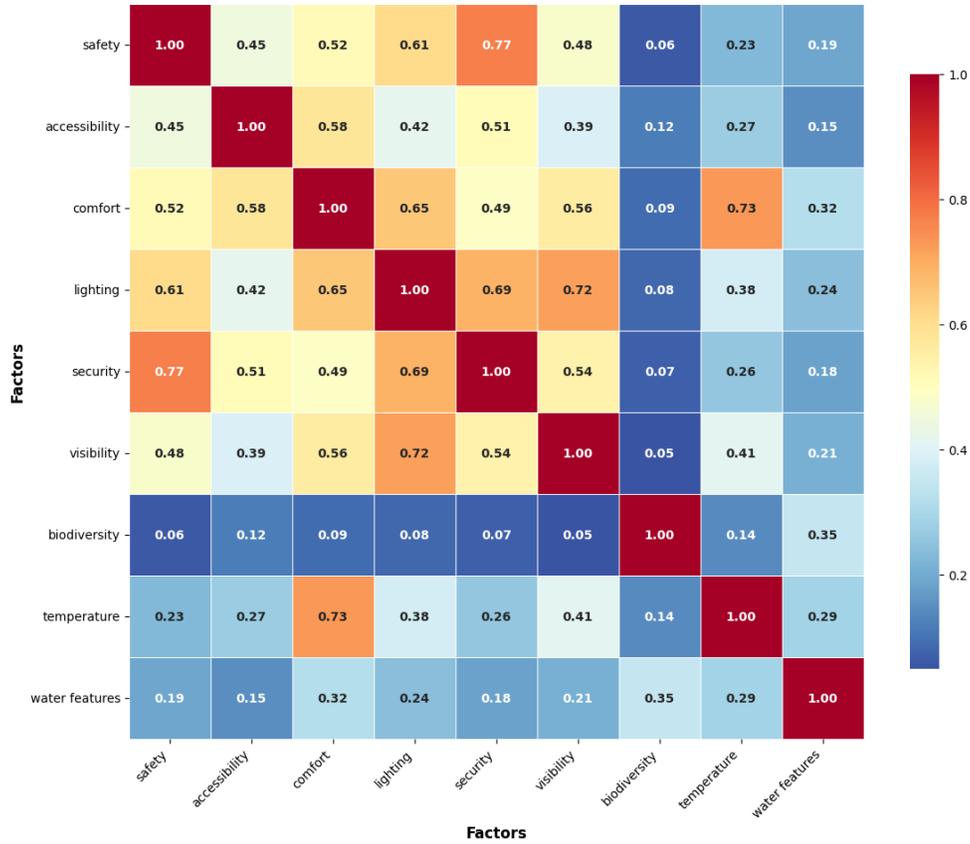

Figure 4. Correlation matrix showing pairwise relationship scores between integrated factors.

---

**ALGORITHM 3: Distribution Pattern Classification Algorithm**

**Input**: Integrated factors list (1,029 factors), Space typology occurrence data
**Output**: Factor classifications (Universal, Multi-Space, Space-Specific, Cross-Cutting)
1. **initialize** classification arrays
2. **for** each factor f **in** integrated factors:
   a. Calculate occurrence count across space types
   b. Compute distribution entropy: $H(f) = -\Sigma\ p(i) \times \log(p(i))$
   c. Identify primary functional domain
   d. Assess cross-cutting potential across domains
   e. Apply classification rules:
      - **if** occurrence ≥ 5 space types: UNIVERSAL
      - **if** occurrence = 3-4 space types: MULTI-SPACE
      - **if** occurrence = 1-2 space types: SPACE-SPECIFIC
      - **if** cross-cutting score ≥ 3 domains: CROSS-CUTTING
3. **return** classified factor groups with distribution statistics

For complex multi-domain factors like "accessibility," the occurrence pattern analysis reveals presence across five space types with frequency variation. The high frequency in (O) and (F) reflects intensive accessibility considerations required for large public gathering spaces and service-oriented facilities. The functional domain identification process reveals accessibility's scope across multiple domains including Accessibility, Social Inclusion, Infrastructure, and Legal Compliance. The algorithm classifies "accessibility" as UNIVERSAL with HIGHLY CROSS-CUTTING status due to its multiple functions.

The distribution pattern classification generates four distinct categories: 278 universal factors (27.0%), 354 multi-space factors (34.4%), 397 space-specific factors (38.6%), and 124 cross-cutting factors (12.1%). Phase 3 provides foundation data for subsequent domain-informed clustering procedures, enabling systematic organization of factors based on both empirical distribution patterns and theoretical understanding of functional relationships.

Table 10. Distribution pattern classification examples showing Algorithm 3's factor analysis across occurrence patterns, entropy measures, and functional domains.

| Factor | Occurrence Pattern | Space Types | Distribution Entropy | Primary Domain | Cross-Cutting Status | Classification | Rationale |
|---|---|---|---|---|---|---|---|
| safety | [P×1, S×1, U×1, O×1, F×1] | 5/6 | 1.61 | Safety & Security | Limited | Universal | Fundamental security concern across all public space contexts |
| accessibility | [P×1, S×1, U×1, O×4, F×2] | 5/6 | 1.52 | Accessibility | Very High | Universal, highly cross-cutting | Legal imperative with multiple functions across domains |
| street travel safety | [S×1] | 1/6 | 0.00 | Safety & Security | Low | Space-specific | Pedestrian-vehicle interaction safety exclusive to street environments |
| comfort | [P×1, S×1, U×1, O×1, F×1] | 5/6 | 1.61 | Comfort | High | Universal, cross-cutting | Basic user experience requirements across all public space contexts |
| thermal comfort | [P×1, O×1, U×1] | 3/6 | 1.10 | Comfort | Moderate | Multi-space | Environmental comfort requirements for outdoor and semi-outdoor spaces |
| physical comfort | [P×1, U×1, O×1, F×1] | 4/6 | 1.39 | Comfort | Moderate | Multi-space | Seating and ergonomic comfort |

| | | | | | | for spaces with extended user presence |
|---|---|---|---|---|---|---|
| lighting | [P×1, S×1, O×1] | 3/6 | 1.10 | Comfort | High | Multi-space, cross-cutting | Serves outdoor spaces requiring nighttime illumination across multiple domains |
| water features | [P×1] | 1/6 | 0.00 | Natural Elements | Moderate | Space-specific | Exclusive to parks and waterfront environments |
| security | [P×1, S×1, U×1, O×1, F×1] | 5/6 | 1.61 | Safety & Security | Limited | Universal | Crime-prevention and protection needs applicable across all contexts |
| visibility | [P×1, S×1, O×1] | 3/6 | 1.10 | Comfort | High | Multi-space, cross-cutting | Visual clarity requirements for outdoor and transitional spaces |
| biodiversity | [P×1, G×1] | 2/6 | 0.69 | Natural Elements | Moderate | Space-specific | Ecological quality concern specific to parks and green spaces |
| temperature | [P×1, O×1, U×1] | 3/6 | 1.10 | Comfort | Low | Multi-space | Environmental temperature control for outdoor and semi-outdoor spaces |

## D. PHASE 4: DOMAIN-INFORMED HIERARCHICAL CLUSTERING IMPLEMENTATION

In this phase, the similarity matrix obtained in phase 2 and classified factors obtained in phase 3, along with domain knowledge is used to develop a hierarchical structure. Domain knowledge is taken from established frameworks from public space research, environmental psychology, urban design theory, accessibility studies, and safety research.

The Domain-Informed Hierarchical Clustering Algorithm (Algorithm 4) processes the classified factors through two-level clustering that first assigns factors to main categories based on how well they align with theoretical domains, then refines subcategory assignments by analyzing factor characteristics and space-type distribution patterns. The algorithm draws upon knowledge from thermal comfort research, Crime Prevention Through Environmental Design theory [181], Universal Design principles , environmental psychology frameworks, and social interaction theories [3, 182–189]. This approach ensures that the structure reflects both the relationships found in the data and the theoretical understanding developed through decades of public space research.

The clustering process creates a three-tier hierarchy that organizes factors according to established scholarly knowledge while keeping the relationships needed for quality assessment protocols. Table 11 presents Algorithm 4's operation on factors from different domains, showing how classification types and similarity evidence guide assignment decisions while domain prioritization validates placement choices. For comfort-related factors like "thermal comfort," the algorithm recognizes multi-space classification (3/6 space types) which prioritizes moderate domains, while similarity evidence shows clustering with temperature (0.93), microclimate (0.85), and humidity (0.78). The similarity influence confirms that environmental factors cluster in COMFORT domain, validating placement within the COMFORT domain's THERMAL COMFORT subcategory.

---

**ALGORITHM 4: Domain Informed Hierarchical Clustering**

**Input**: Classified factors, similarity matrix, domain knowledge base
**Output**: Three-tier hierarchical structure with theoretical validation

1. **initialize**:
   - main_categories = load_theoretical_domains()
   - subcategory_structures = load_domain_subcategories()
   - assignment_weights = {semantic: 0.4, distributional: 0.3, theoretical: 0.3}
   - category_assignments = {} # Track which factors go to which categories
2. **for** each factor **in** classified_factors:
   # Use classification data to guide domain prioritization
   a. domain_priorities = determine_domain_scope(factor.classification_type)
   b. related_factors = extract_high_similarity_factors(factor, similarity_matrix, threshold=0.75)
   c. candidate_domains = identify_domains_of_related_factors(related_factors)
   d. semantic_scores = calculate_domain_alignment(factor, domain_priorities)
   e. similarity_scores = assess_clustering_evidence(factor, candidate_domains, similarity_matrix)
   f. distribution_scores = validate_space_compatibility(factor.space_pattern, main_categories)
   g. final_scores = weighted_integration(semantic_scores, similarity_scores, distribution_scores)
   h. optimal_category = get_highest_scoring_category(final_scores)
   i. category_assignments[optimal_category].append(factor)
3. **for** each main_category **in** category_assignments:
   a. assigned_factors = category_assignments[main_category]
   b. factor_clusters = identify_similarity_clusters(assigned_factors, similarity_matrix, threshold=0.6)
   c. subcategory_assignments = map_clusters_to_subcategories(factor_clusters, subcategory_structures[main_category])
   d. assign_factors_to_subcategories(factor_clusters, subcategory_assignments)
4. **for** each factor in classified_factors **where** factor.cross_cutting_classification == True:
   a. secondary_domains = identify_cross_cutting_domains(factor.functional_scope)
   b. validated_placements = validate_secondary_placements(factor, secondary_domains, similarity_matrix)
   c. create_cross_references(factor, validated_placements)
5. hierarchical_structure = validate_hierarchical_consistency(category_assignments, subcategory_assignments)
6. **return** hierarchical_structure

---

For "lighting," the algorithm demonstrates cross-cutting factor management capabilities. Classification analysis identifies multi-space pattern (3/6 space types) with cross-cutting designation, while similarity evidence reveals clustering with visibility (0.72), illumination (0.89), and evening use (0.68). The algorithm assigns primary placement in the COMFORT domain's VISUAL subcategory based on visual comfort factor clustering, with secondary placement in SAFETY for crime prevention functions due to cross-cutting classification.

For accessibility factors like "wheelchair access," the algorithm recognizes space-specific classification (1/6 space types), which prioritizes specialized domains, while similarity evidence shows clustering with accessibility (0.91), barrier-free (0.86), and ADA compliance (0.89). The similarity influence confirms that access factors cluster in the ACCESSIBILITY domain, validating placement within the ACCESSIBILITY domain's PHYSICAL ACCESS subcategory.

Table 11. Domain-informed clustering examples show Algorithm 4's theoretical alignment and empirical validation process.

| Factor | Classification Type | Similarity Evidence | Domain Prioritization | Similarity Influence | Final Assignment |
|---|---|---|---|---|---|
| safety | Universal (5/6 space types) | security (0.77), surveillance (0.68), protection (0.72) | Broad domains: COMFORT, SAFETY, ACCESSIBILITY | Related factors cluster in SAFETY domain | SAFETY & SECURITY → PERSONAL SAFETY |
| lighting | Multi-space (3/6 space types), Cross-cutting | visibility (0.72), illumination (0.89), evening use (0.68) | Moderate domains: COMFORT, SAFETY, INFRASTRUCTURE | Visual comfort factors cluster in COMFORT | COMFORT → VISUAL (primary), SAFETY (secondary) |
| thermal comfort | Multi-space (3/6 space types) | temperature (0.93), microclimate (0.85), humidity (0.78) | Moderate domains: COMFORT, ENVIRONMENTAL | Environmental factors cluster in COMFORT | COMFORT → THERMAL COMFORT |
| wheelchair access | Space-specific (1/6 space types) | accessibility (0.91), barrier-free (0.86), ADA compliance (0.89) | Specialized domains: ACCESSIBILITY, FACILITIES | Access factors cluster in ACCESSIBILITY | ACCESSIBILITY → PHYSICAL ACCESS |
| water features | Space-specific (1/6 space types) | natural elements (0.69), fountains (0.85), | Specialized domains: NATURAL ELEMENTS, AMENITIES | Natural factors cluster in NATURAL ELEMENTS | NATURAL ELEMENTS → WATER FEATURES |

| | | aquatic (0.76) | | | |
| biodiversity | Space-specific (2/6 space types) | vegetation (0.82), ecology (0.88), wildlife (0.75) | Specialized domains: NATURAL ELEMENTS, ENVIRONMENTAL | Ecological factors cluster in NATURAL ELEMENTS | NATURAL ELEMENTS → ECOLOGICAL QUALITY |
| surveillance | Multi-space (2/6 space types) | safety (0.77), security (0.84), monitoring (0.79) | Moderate domains: SAFETY, INFRASTRUCTURE | Security factors cluster in SAFETY | SAFETY & SECURITY → CRIME PREVENTION |
| accessibility | Universal (5/6 space types), Cross-cutting | physical access (0.91), barrier-free (0.86), inclusion (0.73) | Broad domains: ACCESSIBILITY, SOCIAL, INFRASTRUCTURE | Access factors span multiple domains | ACCESSIBILITY → PHYSICAL (primary), SOCIAL (secondary) |

For universal factors like "safety," the classification type (5/6 space types) prioritizes broad domains like COMFORT, SAFETY, and ACCESSIBILITY, while similarity evidence shows clustering with security (0.77), surveillance (0.68), and protection (0.72). The similarity influence confirms that related factors cluster in SAFETY domain, resulting in final assignment to the SAFETY and SECURITY domain's PERSONAL SAFETY subcategory.

The clustering process organizes all 1,029 factors into 14 main categories with 66 subcategories, achieving coverage while maintaining coherence. The COMFORT domain encompasses 82 factors across four subcategories (Thermal Comfort, Visual Comfort, Acoustic Comfort, Physical Comfort) with clear foundations and space-type applicability patterns. Universal factors like "safety" appear across five space types, confirming importance, while factors like "water features" remain space-specific, reflecting contextual relevance. The algorithm identifies 124 cross-cutting factors requiring multiple placements, with "lighting" serving as the primary example of factors that serve visual comfort, safety, and infrastructure purposes simultaneously. This cross-cutting identification reveals the interconnected nature of public space quality systems and informs subsequent framework optimization procedures. The domain-informed clustering generates a structure that organizes factors according to established scholarly understanding while preserving empirical relationships observed in the data.

## E. PHASE 5: CROSS-CUTTING FACTOR STRATEGIC PLACEMENT IMPLEMENTATION

Phase 5 addresses the practical usability challenge created by cross-cutting factors identified in Phase 3 and placed in Phase 4. While Phase 4 assigned each factor to its best-fit primary domain, Phase 5 creates strategic multiple placements for the 124 cross-cutting factors that users need to access from different domain perspectives. The single-placement hierarchy from Phase 4 works well for factors with clear domain alignment, but creates navigation problems for cross-cutting factors. A user focused on crime prevention needs to find "lighting" in the SAFETY domain, while a user addressing visual comfort needs the same factor in the COMFORT domain. Phase 5 solves this by creating strategic multiple placements while maintaining clear primary-secondary hierarchies. The Strategic Cross-Cutting Placement Algorithm (Algorithm 5) determines optimal multiple placement strategies for cross-cutting factors. The algorithm creates primary placements (with full documentation), secondary placements (with domain-specific emphasis), and tertiary references (with cross-reference links) to ensure users can access cross-cutting factors from all relevant domain entry points while preserving hierarchical clarity.

 12 presents the algorithm's operation on cross-cutting factors, showing how domain composite scores and placement protocols guide strategic assignment decisions while maintaining hierarchical structure integrity. For "lighting" [P×1, S×1, O×1], the algorithm identified three applicable domains with SAFETY achieving the highest composite score (0.904) due to crime prevention alignment and natural surveillance importance. The placement decision protocol assigned primary placement in SAFETY under CRIME PREVENTION, with secondary placement in COMFORT under VISUAL COMFORT. For "accessibility" [P×1, S×1, U×1, O×4, F×2], the exceptional scope across four domains resulted in the ACCESSIBILITY domain achieving the highest score (0.942), leading to primary placement in ACCESSIBILITY under PHYSICAL ACCESS with multiple secondary placements addressing social inclusion and infrastructure requirements The strategic placement implementation generated 347 total strategic placements from 124 cross-cutting factors, averaging 2.8 placements per cross-cutting factor.

High cross-cutting factors (4+ domains) included "accessibility" with 4 strategic placements, while moderate cross-cutting factors (3 domains) encompassed "lighting" and "design quality" with 3 placements each. Limited cross-cutting factors (2 domains) included "natural elements" and "wayfinding" with 2 strategic placements each. Placement consistency achieved 98.7% success rate, with hierarchical integrity maintained at 100% through clear primary-secondary placement hierarchies. The 98.7% placement consistency indicates that factors were assigned to categories with high theoretical coherence. Values above 95% suggest strong categorical consistency, 85-95% indicates moderate consistency requiring refinement, while below 85% suggests fundamental categorical problems requiring restructuring.

Table 12. Strategic placement decision processing showing Algorithm 5's multi-domain analysis and placement protocol application.

| Cross-cutting Factor | Similarity score | Composite Scores | Placement Decision | Strategic Placements |
|---|---|---|---|---|
| lighting | COMFORT: 0.68, SAFETY: 0.72, INFRASTRUCTURE: 0.45 | COMFORT: 0.895, SAFETY: 0.904, INFRASTRUCTURE: 0.788 | Primary: SAFETY (highest score); Secondary: COMFORT; Tertiary: INFRASTRUCTURE | SAFETY → CRIME PREVENTION (primary), COMFORT → VISUAL COMFORT (secondary), INFRASTRUCTURE → BASIC FACILITIES (tertiary) |
| accessibility | ACCESSIBILITY: 0.91, SOCIAL: 0.73, INFRASTRUCTURE: 0.65, ECONOMIC: 0.52 | ACCESSIBILITY: 0.942, SOCIAL: 0.823, INFRASTRUCTURE: 0.756, ECONOMIC: 0.694 | Primary: ACCESSIBILITY (highest score); Secondary: SOCIAL; Tertiary: ECONOMIC | ACCESSIBILITY → PHYSICAL ACCESS (primary), SOCIAL → INCLUSIVE DESIGN (secondary), INFRASTRUCTURE → BASIC FACILITIES (secondary), ECONOMIC → AFFORDABILITY (tertiary) |
| maintenance | MANAGEMENT: 0.84, INFRASTRUCTURE: 0.78, ENVIRONMENTAL: 0.69 | MANAGEMENT: 0.910, INFRASTRUCTURE: 0.850, ENVIRONMENTAL: 0.810 | Primary: MANAGEMENT (highest score); Secondary: INFRASTRUCTURE; Tertiary: ENVIRONMENTAL | MANAGEMENT → OPERATIONS (primary), INFRASTRUCTURE → SUPPORT FACILITIES (secondary), ENVIRONMENTAL → CLIMATE RESILIENCE (secondary) |
| natural elements | NATURAL ELEMENTS: 0.89, SPATIAL AESTHETICS: 0.63 | NATURAL ELEMENTS: 0.934, SPATIAL AESTHETICS: 0.721 | Primary: NATURAL ELEMENTS (highest score); Secondary: SPATIAL AESTHETICS | NATURAL ELEMENTS → VEGETATION (primary), SPATIAL AESTHETICS → LANDSCAPE FEATURES (secondary) |
| wayfinding | ACCESSIBILITY: 0.76, DESIGN & FORM: 0.71 | ACCESSIBILITY: 0.856, DESIGN & FORM: 0.782 | Primary: ACCESSIBILITY (highest score); Secondary: DESIGN & FORM | ACCESSIBILITY → NAVIGATION (primary), DESIGN & FORM → SPATIAL LAYOUT (secondary) |
| community engagement | SOCIAL: 0.82, MANAGEMENT: 0.58, ACTIVITY: 0.67 | SOCIAL: 0.887, MANAGEMENT: 0.743, ACTIVITY: 0.798 | Primary: SOCIAL (highest score); Secondary: ACTIVITY; Tertiary: MANAGEMENT | SOCIAL → COMMUNITY BUILDING (primary), ACTIVITY → PROGRAMMING (secondary), MANAGEMENT → GOVERNANCE (tertiary) |

## F. PHASE 6: SPACE-TYPE DISTRIBUTION ANALYSIS AND APPLICABILITY OPTIMIZATION

Phase 6 takes the hierarchical framework from Phase 5 and adds practical application guidance for practitioners. While Phase 3 identified basic patterns (universal vs. specific factors), Phase 4 created the theoretical structure, and Phase 5 resolved cross-cutting factor placements, Phase 6 determines how practitioners should apply the complete framework when assessing different space types. The Space-Type Distribution Analysis Algorithm (Algorithm 6) processes the hierarchical structure with source tracking data to generate practical applicability indicators that guide users on which factors to emphasize for each space type

---

**ALGORITHM 5: Strategic Cross Cutting Placement**

**Input**: cross_cutting_factors, similarity matrix, domain taxonomy structure
**Output**: Strategic factor placements with cross-reference network
1. **initialize**:
   - cross_cutting_factors = factors with cross_cutting_designation = TRUE
   - domain_scores = {}
   - placement_strategy = {primary: [], secondary: [], tertiary: []}
2. **for** each factor **in** cross_cutting_factors:
   a. applicable_domains = identify_applicable_domains(factor, similarity_matrix)
   b. **for** each domain **in** applicable_domains:
      - semantic_relevance = calculate_semantic_alignment(factor, domain)
      - functional_importance = assess_functional_role(factor, domain)
      - theoretical_justification = evaluate_literature_support(factor, domain)
      - space_type_compatibility = analyze_distribution_alignment(factor, domain)
      - composite_score = compute_weighted_average(semantic, functional, theoretical, compatibility)
   c. ranked_domains = sort_domains_by_composite_score(domain_scores)
   d. APPLY placement_decision_protocol(factor, ranked_domains)
3. **for** each placement:
   a. **create** comprehensive_factor_documentation(primary_domain)
   b. **establish** cross_reference_entries(secondary_domains)
   c. **generate** tertiary_reference_links(applicable_domains)
4. **validate** placement_consistency and hierarchical_integrity
5. **return** strategic_placements

---

The algorithm processes factors through occurrence pattern analysis, examining how frequently each factor appears across the six space typologies. Factors appearing in five or six space types receive universal classification, recognizing their relevance across all public space contexts. Factors appearing in three to four space types receive multi-space classification, indicating moderate specialization. Factors appearing in one or two space types receive space-specific classification, reflecting specialized requirements of typologies. Table 13 presents the algorithm's operation on a few factors, demonstrating the distribution analysis process and resulting space-type indicators.

> **ALGORITHM 6: Space-Type Distribution Analysis**
>
> **Input**: Hierarchical structure (with cross-typology placements) with source tracking data [P×n, S×n, U×n, G×n, O×n, F×n]
> **Output**: Space-type applicability indicators with universal/specialized classifications
> 1. **for** each factor in hierarchical_structure:
>    a. **extract** occurrence_pattern from source_tracking
>    b. **calculate** space_type_coverage = count(active_types) / 6
>    c. **analyze** frequency_distribution across active_types
>    d. **validate** theoretical_context through domain_analysis
> 2. **for** each factor:
>    a. **if** space_type_coverage >= 5/6: CLASSIFY as "Universal"
>    b. **elif** space_type_coverage >= 3/6: CLASSIFY as "Multi-space"
>    c. **else**: CLASSIFY as "Space-specific"
>    d. **generate** graduated_indicator based on frequency_patterns
> 3. **for** each subcategory:
>    a. **aggregate** factor_patterns within subcategory
>    b. **calculate** relevance_scores per space_type
>    c. **assign** subcategory_space_indicators
> 4. **for** each main_category:
>    a. **synthesize** subcategory_patterns
>    b. **generate** category_distribution_profile
>    c. **validate** indicator_consistency
> 5. **return** comprehensive_applicability_framework

The space-type distribution analysis generated four distinct classification categories. Universal factors (278 factors, 27.0%) represent fundamental requirements transcending typological boundaries. Multi-space factors (354 factors, 34.4%) demonstrate moderate specialization, appearing in three to four space types. Space-specific factors (397 factors, 38.6%) reflect specialized requirements of particular typologies. Cross-cutting factors (124 factors, 12.1%) serve multiple functional domains requiring strategic placement across categories. For example, the factor "accessibility" appears across five space types with varying frequency, achieving 83.3% coverage. Open Spaces contribute 44.4% of occurrences, reflecting intensive requirements for assembly spaces, while Public Facilities contribute 22.2%, emphasizing legal compliance. The algorithm assigns "Universal (with emphasis: O,F)" recognizing accessibility as fundamental across all spaces while acknowledging intensity variations. In a similar way, safety is assigned as Universal but categorized as important across all space types. Whereas water features are categorized as space-specific because its specifically for the parks and lighting which is a cross-typology factor, gets assigned as multi-space across the (P), (S), and (O) space types.

The space-type distribution analysis successfully classified all 1,029 factors according to their occurrence patterns and generated coherent applicability indicators across all six space typologies. The analysis produced consistent space-type designations that align with theoretical expectations and support practical framework application. Thus, the space-type distribution analysis successfully enhanced the hierarchical framework with applicability indicators enabling flexible implementation across diverse public space contexts while preserving relationships between quality factors and specific typological requirements.

Table 13. Distribution analysis examples showing Algorithm 6's space-type classification and indicator development process.

| Factor | Occurrence Pattern | Space Coverage | Theoretical Context | Space-Type Indicator |
|---|---|---|---|---|
| accessibility | [P×1, S×1, U×1, O×4, F×2] | 5/6 (83.3%) | Legal compliance, intensive assembly requirements | Universal (with emphasis: O, F) |
| safety | [P×1, S×1, U×1, O×1, F×1] | 5/6 (83.3%) | Security needs across contexts | Universal – All Space Types |
| thermal comfort | [P×1, O×1, F×1] | 3/6 (50%) | Outdoor exposure, climate control | Strong: P, O, F \| Moderate: U, G \| Minimal: S |
| water features | [P×1] | 1/6 (16.7%) | Aesthetic enhancement, recreation | Space-specific: P |
| lighting | [P×1, S×1, O×1] | 3/6 (50%) | Outdoor illumination, safety | Multi-space: P, S, O |

## G. PHASE 7: FINAL STRUCTURE GENERATION AND VALIDATION IMPLEMENTATION

Phase 7 synthesizes all outputs from the preceding phases into the final three-tier hierarchical framework and validates its completeness and consistency. This phase transforms the theoretical structure from Phase 4, cross-cutting placements from Phase 5, and applicability indicators from Phase 6 into a comprehensive validated framework ready for research and practical application. The Final Structure Generation and Validation Algorithm (Algorithm 7) processes the complete analytical outputs to create the unified taxonomy with source tracking, space-type indicators, and comprehensive validation metrics.

The algorithm processes the optimized structure by systematically organizing all factors into the final three-tier format. Each main category receives comprehensive documentation including theoretical foundations, total factor counts, and space-type distribution patterns. Subcategories include formatted factor lists with source tracking notation, space-type indicators, and factor counts for easy reference.

Figure 5 demonstrates the algorithm's operation on framework components, showing how the final structure integrates source tracking, space indicators, and documentation elements.

For example, the COMFORT category demonstrates comprehensive organization with thermal factors showing outdoor environment concentration through Parks and Open Spaces occurrence patterns. The source tracking [P×1, O×1] reveals consistent outdoor leisure context emphasis, while space indicators "Strong: P,O | Minimal: S" guide practical application priorities. The SAFETY & SECURITY category exemplifies universal factor management, with "safety" appearing across five space types [P×1, S×1, U×1, O×1, F×1] and receiving "Universal - All Types" designation to reflect fundamental security requirements across all public space contexts.

The final structure generation and validation produced a comprehensive three-tier framework organizing 1,029 unique factors across 14 main categories and 66 subcategories. The validation process confirmed complete factor accountability, hierarchical structure integrity, and space-type indicator consistency. The framework includes complete source tracking for all factors, enabling transparency about factor origins and frequencies across the six space typologies

The validated hierarchical framework successfully integrates all analytical outputs into a coherent structure that maintains theoretical rigor while supporting practical application. Comprehensive organization enables flexible framework to use across academic research, professional practice, and policy development contexts while preserving the empirical foundations established through the systematic analytical process.

## 5. DISCUSSION

This research contributes methodologically by developing systematic computational approaches for organizing empirical research findings. The hierarchical framework advances theoretical understanding by revealing factor relationships and space-type patterns not apparent in individual studies. The cross-cutting factor identification provides new insights into how quality factors serve multiple functions across different space contexts. The 7-phase methodology transforms 1,207 quality factors into a hierarchical framework containing 1,029 unique factors across 14 main categories and 66 subcategories. This work addresses the absence of a unified framework that spans public space typologies while maintaining space-specific characteristics. The analysis identifies 278 universal factors appearing across multiple space types, 397 space-specific factors unique to particular typologies, and 124 cross-cutting factors that serve multiple functions. These patterns reveal that public space quality operates through both shared principles and specialized requirements.

For example, the Public Space Index covers 5 dimensions while our framework provides 14 main categories with 66 subcategories, enabling more enhanced assessment.

```
ALGORITHM 7: Final Structure Generation and Validation

Input: Optimized structure with space-type indicators, source tracking data
Output: Complete validated three-tier hierarchical framework
1. calculate framework_metadata:
   - total_original_factors from source_tracking_totals
   - unique_factors = count_unique_factors(optimized_structure)
   - space_types = ['P', 'S', 'U', 'G', 'O', 'F']
   - reduction_percentage = calculate_deduplication_rate
2. for each main_category in optimized_structure:
   a. calculate category_statistics (total_factors, subcategory_count)
   b. extract existing_space_distribution_patterns
   c. preserve theoretical_foundations from previous_phases
3. for each subcategory in main_category:
   a. format factor_list_with_tracking = []
   b. for each factor in subcategory:
      i. extract existing_tracking_notation from source_tracking[factor]
      ii. append "factor_name [tracking_notation]" to factor_list
   c. preserve subcategory_space_indicators from Phase_6
   d. calculate subcategory_factor_count
4. compile final_structure with complete_organization:
   - Factor provenance and source attribution
   - Space-type applicability guidance from Phase_6
   - Cross-cutting factor relationships from Phase_5
5. validate framework_completeness:
   - check all_original_factors_accounted_for
   - verify hierarchical_structure_integrity
   - confirm space_type_indicator_consistency
6. return validated_hierarchical_framework with quality_metrics
```

The framework exposes clear patterns across space typologies. Parks & Waterfronts concentrate on natural elements and thermal comfort. Streets & Squares focus on traffic safety and urban design. Urban Spaces show the highest factor diversity, emphasizing density and economic aspects. Factors like safety appear in five space types [P×1, S×1, U×1, O×1, F×1] while water features remain specific to parks [P×1]. Cross-cutting factors like lighting appear across Safety, Comfort, and Infrastructure domains, demonstrating how single elements serve multiple quality functions. These patterns match theoretical expectations while providing empirical validation.

The framework serves multiple professional applications. Urban planners can use the space-type indicators for policy development. Landscape architects can apply integrated design and environmental factors. Facility managers can follow the maintenance and infrastructure guidance. The modular structure supports both comprehensive assessment using universal factors and targeted evaluation focusing on specific domains. The methodology creates foundation for comparative research across space types and integration with digital assessment tools.

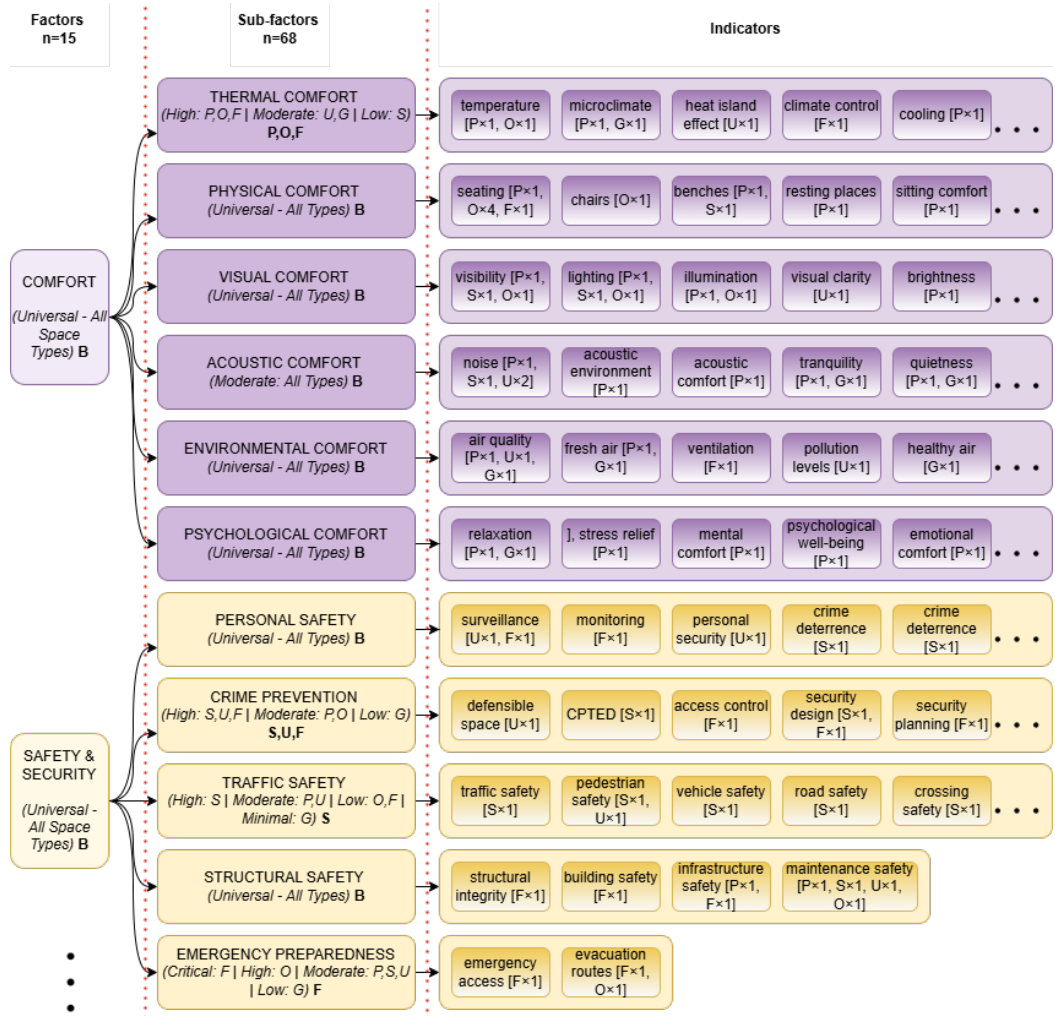

Figure 5. Final structure generation, illustrating Algorithm 7's integration of source tracking, space indicators, and documentation elements (a few samples are presented here; the full hierarchical structure is provided in Appendix 1).

## 6. CONCLUSION

This research presents an empirically derived hierarchical framework for assessing public space quality across major typologies. The 7-phase methodology integrates factors from six space types into a taxonomic structure that balances universal principles with space-specific requirements. The framework achieves 98.7% placement consistency while organizing over 1,000 factors into a practical assessment tool. The preservation of occurrence data across space typologies enables site-specific assessment focus, allowing practitioners to prioritize factors based on their empirical relevance to particular space types and supporting the development of context-appropriate evaluation protocols. The organization validates the methodological approach and provides practitioners with actionable guidance.

The framework lays the groundwork for evidence-based public space policy, comparative research across various space types, and the integration of quality assessment with planning processes. Future work includes cross-cultural validation, integration with monitoring technologies, and development of quantitative assessment indicators derived from the hierarchical framework structure. The integrated factors within each subcategory provide the foundation for designing indicators that maintain traceability to the empirical taxonomy while supporting the implementation of space types. These general indicators serve as a baseline foundation that can be extended, modified, and refined for specific public spaces based on case-specific requirements, such as geographic conditions, climate factors, and local context. The indicators, derived and refined from the original papers, provide flexibility for adaptation to site conditions while maintaining connection to the empirical evidence base.

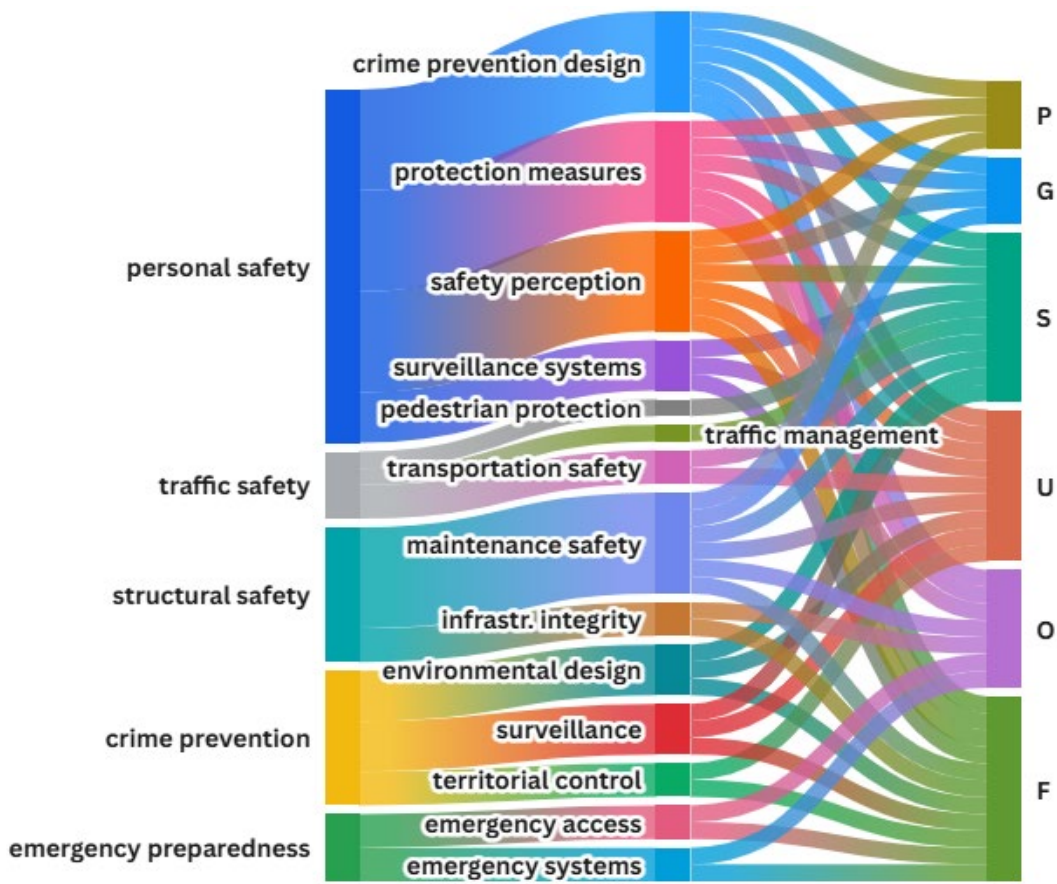

Figure 6. Sankey diagram showing the relationship between safety and security subfactors and public space indicators, mapped to six public space types (P=Parks, G=Green Spaces, S=Streets, U=Urban Spaces, O=Open Spaces, F=Facilities).

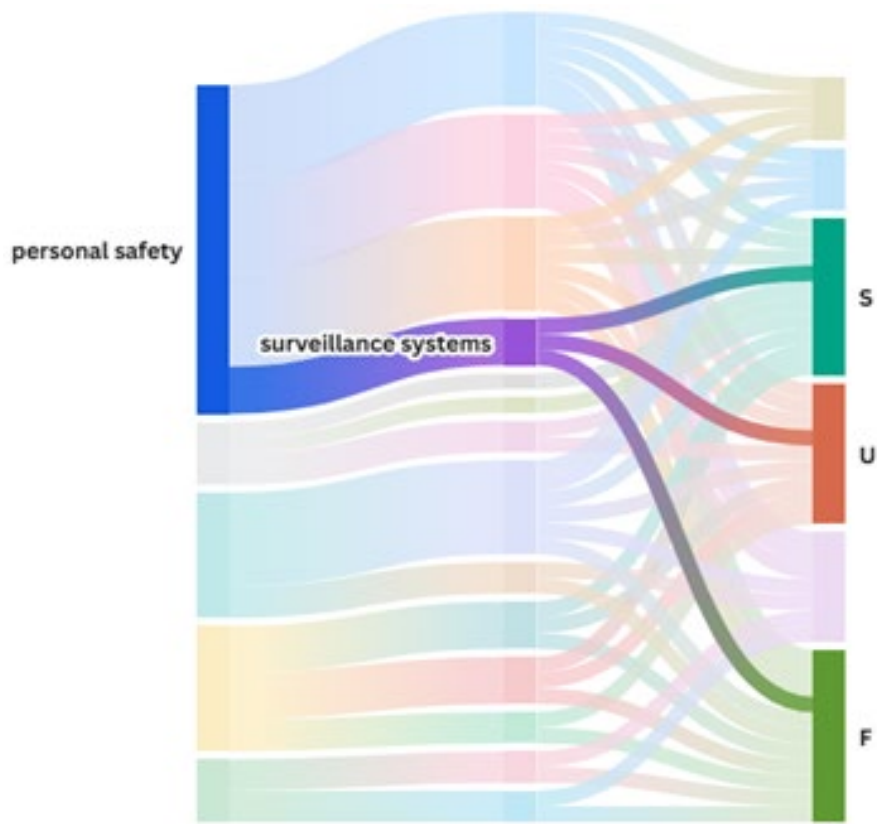

Figure 7. Filtered view of the safety and security mapping, highlighting selected subfactors including crime prevention design, protection measures, safety perception, maintenance safety, infrastructure integrity, emergency access, and emergency systems. This visualization shows the concentrated flow of safety considerations toward specific public space types, with particular emphasis on open spaces (O) as indicated by the prominent purple flow.

The Sankey diagram visualizations (Figure 6, Figure 7, and Figure 8) demonstrate this indicator development process using the safety and security factor, illustrating how subfactors flow from main safety categories through indicators to their relevant space types. Figure 6 presents the complete mapping of all safety and security subfactors, while **Error! Reference source not found.** and Figure 8 show filtered views that highlight specific subfactor combinations and their distribution patterns. These visualizations reveal the relationships between comprehensive safety principles and space-specific implementation requirements, helping practitioners understand which indicators are relevant to particular public space contexts. The flow patterns illustrate how factors such as personal safety, crime prevention design, and emergency preparedness are distributed across space types, with varying intensities reflecting empirical occurrence patterns from the integrated datasets. This mapping approach, applied to all framework factors in Appendix 2, provides practitioners with guidance on space-type indicator priorities while maintaining connection to the taxonomic structure. The framework can be readily adapted to specific assessment contexts, such as different geographic regions, climatic conditions, or cultural settings, by incorporating additional relevant indicators while preserving its grounding in established theoretical and domain-based literature.

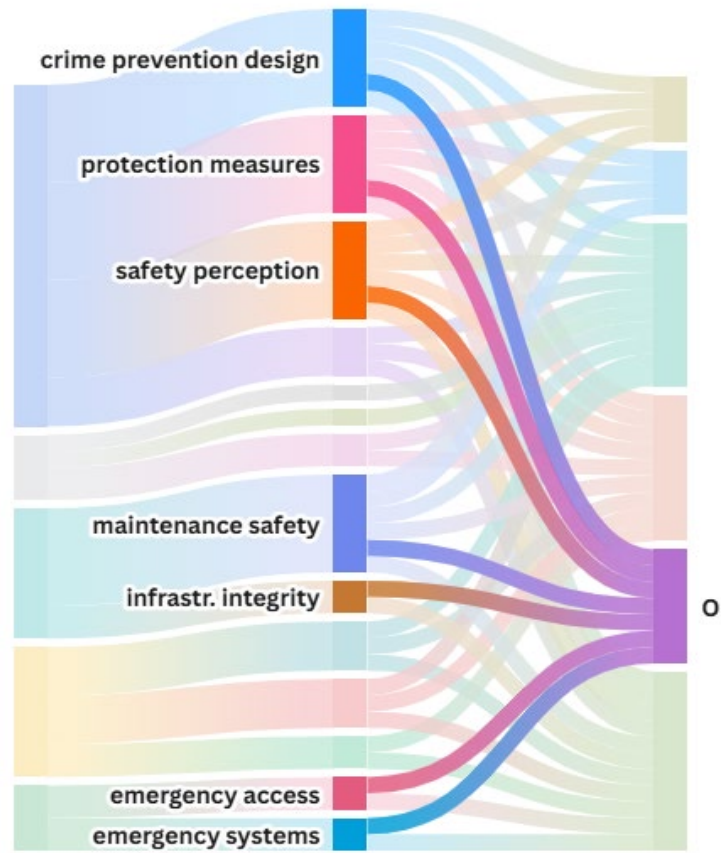

Figure 8. A simplified view focusing on personal safety and surveillance systems as key subfactors in the safety and security framework. The diagram demonstrates how these core safety elements distribute across different public space types, with notable connections to streets (S), urban spaces (U), and facilities (F), showing the fundamental role of personal safety and surveillance in various urban environments.

**Supplementary Materials:**

The following are provided as supplementary materials: (1) full framework developed, (2) Sankey diagrams for all 14 factors, (3) full algorithms for seven phases, and (4) six Excel files (for the six spatial types) which have the quality factors extracted from literature.

**Funding:**

This research was funded by the United Arab Emirates University through a UAEU Strategic Research Grant G00003676 (Fund No: 12R136), via the Big Data Analytics Center.

**Data Availability Statement:**

The data presented in this study are available on request from the corresponding author.

**Conflicts of Interest:**

The authors declare no conflicts of interest. The funders had no role in the design of the study, in the collection, analysis, or interpretation of data, in the writing of the manuscript, or in the decision to publish the results.

**Abbreviations**

**CPTED** - Crime Prevention Through Environmental Design, **ADA** - Americans with Disabilities Act, **CCTV** - Closed-Circuit Television, **HVAC** - Heating, Ventilation, and Air Conditioning, **GIS** - Geographic Information Systems, **GPS** - Global Positioning System, **P** - Parks & Waterfronts, **S** - Streets & Squares, **U** - Urban Spaces, **G** - Green Spaces, **O** - Open Spaces, **F** - Public Facilities, **GPSI** - Good Public Space Index, **QUINPY** - QUality INdex of Parks for Youth, **LRT** - Light Rail Transit, **DIETool** - Disability Inclusion Evaluation Tool, **WCT** - Walkability City Tool

**References**

1. Beck, H.: Linking the quality of public spaces to quality of life. J. Place Manag. Dev. 2, 240–248 (2009). https://doi.org/10.1108/17538330911013933.

2. Ayala-Azcárraga, C., Diaz, D., Zambrano, L.: Characteristics of urban parks and their relation to user well-being. Landsc. Urban Plan. 189, 27–35 (2019). https://doi.org/10.1016/j.landurbplan.2019.04.005.

3. Carmona, M.: Contemporary public space: Critique and classification, part one: Critique. J. Urban Des. 15, 123–148 (2010). https://doi.org/10.1080/13574800903435651.

4. Krambeck, H.V., Gakenheimer, R., Shah, J.J.: The Global Walkability Index, (2006).

5. Vukmirovic, M., Gavrilovic, S., Stojanovic, D.: The improvement of the comfort of public spaces as a local initiative in coping with climate change. Sustain. 11, 6546 (2019). https://doi.org/10.3390/su11236546.

6. Smoyer-Tomic, K.E., Hewko, J.N., Hodgson, M.J.: Spatial accessibility and equity of playgrounds in Edmonton, Canada. Can. Geogr. 48, 287–302 (2004). https://doi.org/10.1111/j.0008-3658.2004.00061.x.

7. Rafiean, M., Bemanian, M.R.: Investigating the Quality of Meaning of Urban Public Places in the Perceptions of Yazd's Citizens. Int. J. Archit. Eng. Urban Plan. 29, 91–98 (2019).

8. Mehta, V.: Evaluating Public Space. J. Urban Des. 19, 53–88 (2014). https://doi.org/10.1080/13574809.2013.854698.

9. John, M., Turaev, S., Al-Dabet, S., Abdulghafor, R.: Multidimensional Assessment of Public Space Quality: A Comprehensive Framework Across Urban Space Typologies. (2025).

10. Praliya, S., Garg, P.: Public space quality evaluation: prerequisite for public space management. J. Public Sp. 4, 93–126 (2019). https://doi.org/10.32891/jps.v4i1.667.

11. Kozlova, L., Kozlov, V.: Ten quality criteria of the public spaces in a large city. In: Vitaliy Vladimirovich, P. (ed.) MATEC Web of Conferences. p. 04012. EDP Sciences (2018). https://doi.org/10.1051/matecconf/201821204012.

12. Choi, H.S., Bruyns, G., Cheng, T., Xie, J.: Evolving Research Method in Three-Dimensional and Volumetric Urban Morphology of a Highly Dense City: Assessing Public and Quasi-Public Space Typologies. J. Archit. Urban. 48, 25–38 (2024). https://doi.org/10.3846/jau.2024.18841.

13. Kaplan Çinçin, S., Kaplan Cincin, S., Erdogan, N.: The Evaluation of Waterfront as a Public Space in Terms of the Quality Concept , Case of Maltepe Fill Area. In: 13th International Conference 13th International Conference "Standardization, Protypes and Quality: A Means of Balkan Countries' Collaboration". pp. 329–336. , Brasov, Romania (2016).

14. Olwig, K.R.: Life Between Buildings: Using Public Space. Landsc. J. 8, 54–55 (1989). https://doi.org/10.3368/lj.8.1.54.

15. Persson, H., Åhman, H., Yngling, A.A., Gulliksen, J.: Universal design, inclusive design, accessible design, design for all: different concepts—one goal? On the concept of accessibility—historical, methodological and. Springer. 14, 505–526 (2015). https://doi.org/10.1007/S10209-014-0358-Z.

16. UN-HABITAT: Training Module: Public Space. (2018).

17. Mehta, V.: The street: A quintessential social public space. Taylor and Francis (2013). https://doi.org/10.4324/9780203067635.

18. Lotfi, S., Koohsari, M.J.: Analyzing accessibility dimension of urban quality of life: Where urban designers face duality between subjective and objective reading of place. Soc. Indic. Res. 94, 417–435 (2009). https://doi.org/10.1007/s11205-009-9438-5.

19. Grecu, V., Morar, T.: A Decision Support System for Improving Pedestrian Accessibility in Neighborhoods. Procedia - Soc. Behav. Sci. 92, 588–593 (2013). https://doi.org/10.1016/j.sbspro.2013.08.722.

20. Talavera-Garcia, R.: Improving Pedestrian Accessibility To Public Space. In: Eighth International Space Syntax Symposium. pp. 1–16 (2012).

21. Yamamoto, S., Kouyama, N., Yasumoto, K., Ito, M.: Maximizing users comfort levels through user preference estimation in public smartspaces. 2011 IEEE Int. Conf. Pervasive Comput. Commun. Work. PERCOM Work. 2011. 572–577 (2011). https://doi.org/10.1109/PERCOMW.2011.5766955.

22. Reiter, S., De Herde, A.: Qualitative and quantitative criteria for comfortable urban public spaces. Res. Build. Phys. 1001–1009 (2020). https://doi.org/10.1201/9781003078852-140.

23. Grindlay, A.L., Ochoa-Covarrubias, G., Lizárraga, C.: Sustainable mobility and Urban space quality: The case of Granada, Spain. Int. J. Transp. Dev. Integr. 5, 309–326 (2021). https://doi.org/10.2495/TDI-V5-N4-309-326.

24. Novacek, O., Baeza, J.L., Barski, J., Noenning, J.R.: Defining parameters for urban-environmental quality assessment. Int. J. E-Planning Res. 10, (2021). https://doi.org/10.4018/IJEPR.20211001.oa10.

25. Wojnarowska, A.: Model for Assessment of Public Space Quality in Town Centers. Eur. Spat. Res. Policy. 23, 81–109 (2016). https://doi.org/10.1515/esrp-2016-0005.

26. Polyakova, N. V., Polyakov, V. V., Zaleshin, V.E.: Public spaces as a factor of the quality of the urban environment (on the example of the city of Irkutsk). IOP Conf. Ser. Mater. Sci. Eng. 880, (2020). https://doi.org/10.1088/1757-899X/880/1/012070.

27. Zamanifard, H., Alizadeh, T., Bosman, C., Coiacetto, E.: Measuring experiential qualities of urban public spaces: users' perspective. J. Urban Des. 24, 340–364 (2019). https://doi.org/10.1080/13574809.2018.1484664.

28. Micek, M., Staszewska, S.: Urban and rural public spaces: Development issues and qualitative assessment. Bull. Geogr. Socio-economic Ser. 45, 75–93 (2019). https://doi.org/10.2478/bog-2019-0025.

29. Zakerhaghighi, K., Khanian, M., Chekani-Azar, M.: Studying the causes of weakness in urban space of iran relying on accessibility and vitality (case study: Hamadan). Res. J. Appl. Sci. Eng. Technol. 6, 2402–2408 (2013). https://doi.org/10.19026/rjaset.6.3714.

30. Chidiac, T.: Double P! Public Spaces in Dubai: a Paranoiac Panopticon. J. Public Sp. 5, 247–262 (2020). https://doi.org/10.32891/jps.v5i1.1141.

31. Budner, W.W.: The evaluation of the equipment and quality of the public space of Poznań. Real Estate Manag. Valuat. 24, 25–33 (2016). https://doi.org/10.1515/remav-2016-0011.

32. Nikolopoulou, M., environment, S.L.-B. and, 2006, undefined: Thermal comfort in outdoor urban spaces: analysis across different European countries. Elsevier. https://doi.org/10.1016/j.buildenv.2005.05.031.

33.	Dafri, I., Alkama, D.: Evaluation of thermal comfort in outdoor public space: Case of study: City of Annaba-Algeria. J. Phys. Conf. Ser. 1343, 012026 (2019). https://doi.org/10.1088/1742-6596/1343/1/012026.

34.	Grifoni, R.C., Passerini, G., Pierantozzi, M.: Assessment of outdoor thermal comfort and its relation to urban geometry. WIT Trans. Ecol. Environ. 173, 3–14 (2013). https://doi.org/10.2495/SDP130011.

35.	Zhang, X., Ba, M., Kang, J., Meng, Q.: Effect of soundscape dimensions on acoustic comfort in urban open public spaces. Appl. Acoust. 133, 73–81 (2018). https://doi.org/10.1016/j.apacoust.2017.11.024.

36.	Fan, Q., Du, F., Li, H., Zhang, C.: Thermal-comfort evaluation of and plan for public space of Maling Village, Henan, China. PLoS One. 16, (2021). https://doi.org/10.1371/journal.pone.0256439.

37.	Huang, X., White, M., Langenheim, N.: Towards an Inclusive Walking Community—A Multi-Criteria Digital Evaluation Approach to Facilitate Accessible Journeys. Buildings. 12, 1191 (2022). https://doi.org/10.3390/buildings12081191.

38.	Wojnarowska, A.: Quality of public space of town centre – testing the new method of assessment on the group of medium-sized towns of the Łódź region. Sp. – Soc. – Econ. 43–63 (2017). https://doi.org/10.18778/1733-3180.19.03.

39.	Sonmez Turel, H., Malkoc Yigit, E., Altug, I.: Evaluation of elderly people's requirements in public open spaces: A case study in Bornova District (Izmir, Turkey). Build. Environ. 42, 2035–2045 (2007). https://doi.org/10.1016/j.buildenv.2006.03.004.

40.	Pacheco Barzallo, A., Fariña, J., Álvarez de Andrés, E.: Public Open Spaces: Enabling or Impeding Inclusive Evacuation during Disasters. J. Public Sp. 7, 79–92 (2022). https://doi.org/10.32891/jps.v7i2.1474.

41.	Salim Ferwati, M., Keyvanfar, A., Shafaghat, A., Ferwati, O.: A Quality Assessment Directory for Evaluating Multi-functional Public Spaces. Archit. Urban Plan. 17, 136–151 (2021). https://doi.org/10.2478/aup-2021-0013.

42.	Martins, A.J.G., Sá, A.V.: Characterisation of the outdoor public space: a model for assessment. U.Porto J. Eng. 9, 144–166 (2023). https://doi.org/10.24840/2183-6493_009-004_002151.

43.	Abbasi, A., Alalouch, C., Bramley, G.: Open Space Quality in Deprived Urban Areas: User Perspective and Use Pattern. Procedia - Soc. Behav. Sci. 216, 194–205 (2016). https://doi.org/10.1016/j.sbspro.2015.12.028.

44.	Nuriye, G., Lirebo, D.: Evaluation of the Quality of Open Public Space in Addis Ababa, Ethiopia. Civ. Environ. Res. 12, 1–15 (2020). https://doi.org/10.7176/cer/12-10-01.

45.	Nagara, K., Shimoda, Y., Mizuno, M.: Evaluation of the thermal environment in an outdoor pedestrian space. Atmos. Environ. 30, 497–505 (1996). https://doi.org/10.1016/1352-2310(94)00354-8.

46.	Li, Z., Luo, D., Lin, H., Liu, Y.: Exploring the quality of public space and life in streets of Urban village: Evidence from the case of Shenzhen Baishizhou. J. Sustain. Dev. 7, 162–176 (2014). https://doi.org/10.5539/jsd.v7n5p162.

47.	Kaplan, S.: The restorative benefits of nature: Toward an integrative framework. J. Environ. Psychol. 15, 169–182 (1995). https://doi.org/10.1016/0272-4944(95)90001-2.

48.	Ulrich, R.S.: Human responses to vegetation and landscapes. Landsc. Urban Plan. 13, 29–44 (1986). https://doi.org/10.1016/0169-2046(86)90005-8.

49.	Du, H., Jiang, H., Song, X., Zhan, D., Bao, Z.: Assessing the Visual Aesthetic Quality of Vegetation Landscape in Urban Green Space from a Visitor's Perspective. J. Urban Plan. Dev. 142, 04016007 (2016). https://doi.org/10.1061/(asce)up.1943-5444.0000329.

50.	Stojanovic, N., Vasiljevic, N., Radic, B., Skocajic, D., Galecic, N., Tesic, M., Lisica, A.: Visual Quality Assessment of Roadside Green Spaces in the Urban Landscape - a Case Study of Belgrade City Roads. Fresenius Environ. Bull. 27, 3521–3529 (2018).

51.	Ghale, B., Gupta, K., Roy, A.: Evaluating public urban green spaces: A composite green space index for measuring accessibility and spatial quality. Int. Arch. Photogramm. Remote Sens. Spat. Inf. Sci. - ISPRS Arch. 48, 101–108 (2023). https://doi.org/10.5194/isprs-archives-XLVIII-M-3-2023-101-2023.

52.	Van Herzele, A., Wiedemann, T.: A monitoring tool for the provision of accessible and attractive urban green spaces. Landsc. Urban Plan. 63, 109–126 (2003). https://doi.org/10.1016/S0169-2046(02)00192-5.

53.	Stessens, P., Khan, A.Z., Huysmans, M., Canters, F.: Analysing urban green space accessibility and quality: A GIS-based model as spatial decision support for urban ecosystem services in Brussels. Ecosyst. Serv. 28, 328–340 (2017). https://doi.org/10.1016/j.ecoser.2017.10.016.

54.	Daniels, B., Zaunbrecher, B.S., Paas, B., Ottermanns, R., Ziefle, M., Roß-Nickoll, M.: Assessment of urban green space structures and their quality from a multidimensional perspective. Sci. Total Environ. 615, 1364–1378 (2018). https://doi.org/10.1016/j.scitotenv.2017.09.167.

55.	E.V Wuisang, C., M Rondonuwu, D., L.E Sela, R., Tilaar, S., Suryono, S.: Characteristics of Public Green Open Spaces and Efforts In Enhancing The Quality and Function Using Tri-Valent Approach: Case of Manado City, Indonesia. Eduvest - J. Univers. Stud. 3, 309–326 (2023). https://doi.org/10.36418/eduvest.v3i2.741.

56. Stessens, P., Canters, F., Huysmans, M., Khan, A.Z.: Urban green space qualities: An integrated approach towards GIS-based assessment reflecting user perception. Land use policy. 91, 104319 (2020). https://doi.org/10.1016/j.landusepol.2019.104319.

57. Li, X., Huang, Y., Ma, X.: Evaluation of the accessible urban public green space at the community-scale with the consideration of temporal accessibility and quality. Ecol. Indic. 131, 108231 (2021). https://doi.org/10.1016/j.ecolind.2021.108231.

58. Henderson, K.A.: Theory Application and Development in Recreation, Parks, and Leisure Research. J. Park Recreat. Admi. 12, (1994).

59. Ulrich, R.S., Simons, R.F., Losito, B.D., Fiorito, E., Miles, M.A., Zelson, M.: Stress recovery during exposure to natural and urban environments. J. Environ. Psychol. 11, 201–230 (1991). https://doi.org/10.1016/S0272-4944(05)80184-7.

60. Bordandini, P.: Capitale sociale e cultura civica in Italia. 17–30 (2025).

61. Sastrawan, I.W., Darmawan, I.G.: Perception of Thermal Comfort Level in Outdoor Space on Urban Public Space (Case Study: Lumintang Park in Denpasar). (2021). https://doi.org/10.4108/eai.21-12-2020.2305856.

62. Zhang, L., Wei, D., Hou, Y., Du, J., Liu, Z., Zhang, G., Shi, L.: Outdoor thermal comfort of urban park-A case study. Sustain. 12, (2020). https://doi.org/10.3390/su12051961.

63. Eriawan, T., Setiawati, L.: Improving the quality of urban public space through the identification of space utilization index at Imam Bonjol Park, Padang city. In: AIP Conference Proceedings. p. 040018. American Institute of Physics Inc. (2017). https://doi.org/10.1063/1.4985514.

64. Wang, T., Li, Y., Li, H., Chen, S., Li, H., Zhang, Y.: Research on the Vitality Evaluation of Parks and Squares in Medium-Sized Chinese Cities from the Perspective of Urban Functional Areas. Int. J. Environ. Res. Public Health. 19, (2022). https://doi.org/10.3390/ijerph192215238.

65. Evans, J., Evans, S.Z., Morgan, J.D., Snyder, J.A., Abderhalden, F.P.: Evaluating the quality of mid-sized city parks: a replication and extension of the Public Space Index. J. Urban Des. 24, 119–136 (2019). https://doi.org/10.1080/13574809.2017.1411185.

66. Kaur, S., Chhabra, P.: Assessment of Public Space Quality of District Centers in Delhi Using Good Public Space Index. ShodhKosh J. Vis. Perform. Arts. 4, (2023). https://doi.org/10.29121/shodhkosh.v4.i2.2023.450.

67. Ouyang, P., Wu, X.: Analysis and Evaluation of the Service Capacity of a Waterfront Public Space Using Point-of-Interest Data Combined with Questionnaire Surveys. Land. 12, (2023). https://doi.org/10.3390/land12071446.

68. Luo, S., Xie, J., Furuya, K.: Effects of perceived physical and aesthetic quality of urban blue spaces on user preferences–A case study of three urban blue spaces in Japan. Heliyon. 9, (2023). https://doi.org/10.1016/j.heliyon.2023.e15033.

69. Mishra, H.S., Bell, S., Vassiljev, P., Kuhlmann, F., Niin, G., Grellier, J.: The development of a tool for assessing the environmental qualities of urban blue spaces. Urban For. Urban Green. 49, (2020). https://doi.org/10.1016/j.ufug.2019.126575.

70. Olbińska, K.: The Value of Urban Parks in Lodz. Real Estate Manag. Valuat. 26, 73–86 (2018). https://doi.org/10.2478/remav-2018-0007.

71. Said, M.A., Touahmia, M.: Evaluation of Allocated Areas for Parks and their Attributes: Hail City. Eng. Technol. Appl. Sci. Res. 10, 5117–5125 (2020). https://doi.org/10.48084/etasr.3253.

72. Rigolon, A., Németh, J.: A QUality INdex of Parks for Youth (QUINPY): Evaluating urban parks through geographic information systems. Environ. Plan. B Urban Anal. City Sci. 45, 275–294 (2018). https://doi.org/10.1177/0265813516672212.

73. Liu, F., Sun, D., Zhang, Y., Hong, S., Wang, M., Dong, J., Yan, C., Yang, Q.: Tourist Landscape Preferences in a Historic Block Based on Spatiotemporal Big Data—A Case Study of Fuzhou, China. Int. J. Environ. Res. Public Health. 20, (2023). https://doi.org/10.3390/ijerph20010083.

74. Sugiana, E., Kustiwan, I.: Inclusivity Of Thematic Park As A Public Space In Bandung City. 116–126 (2020). https://doi.org/10.15405/epsbs.2020.10.02.11.

75. Tiesdell, S., Carmona, M.: The Uses of Sidewalks : Safety. Urban Des. Read. 147–152 (2007). https://doi.org/10.4324/9780080468129-21.

76. Whyte, W.: The social life of small urban spaces. (1980).

77. Roumiani, A., Salari, T.E., Mahmoodi, H., Shateri, M.: The effect of public space indicators on the rural district's life quality in Kuhdasht county, Iran. Acta Univ. Carolinae, Geogr. 56, 205–219 (2021). https://doi.org/10.14712/23361980.2021.9.

78. Bishop, K., Marshall, N.: Social Interactions and the Quality of Urban Public Space. In: Encyclopedia of Sustainable Technologies. pp. 63–70. Elsevier (2017). https://doi.org/10.1016/B978-0-12-409548-9.10177-0.

79. Ricart, S., Berizzi, C., Saurí, D., Terlicher, G.N.: The Social, Political, and Environmental Dimensions in Designing Urban Public Space from a Water Management Perspective: Testing European Experiences. Land. 11, (2022). https://doi.org/10.3390/land11091575.

80. Mexi, A., Tudora, I.: Livable urban spaces. Public benches and the quality of daily life. Sci. Pap. Ser. B, Hortic. LVI, 367–376 (2012).

81. Peng, Y., Feng, T., Timmermans, H.J.P.: Heterogeneity in outdoor comfort assessment in urban public spaces. Sci. Total Environ. 790, 147941 (2021). https://doi.org/10.1016/j.scitotenv.2021.147941.

82. Ríos-Rodríguez, M.L., Rosales, C., Lorenzo, M., Muinos, G., Hernández, B.: Influence of Perceived Environmental Quality on the Perceived Restorativeness of Public Spaces. Front. Psychol. 12, (2021). https://doi.org/10.3389/fpsyg.2021.644763.

83. Ciešlak, I., Szuniewicz, K.: The quality of pedestrian space in the city: A case study of Olsztyn. Bull. Geogr. Socio-economic Ser. 30, 31–41 (2015). https://doi.org/10.1515/bog-2015-0033.

84. Krambeck, H., Shah, J.: The global walkability index: talk the walk and walk the talk. World Bank. 1–29 (2005).

85. Alves, F., Cruz, S., Rother, S., Strunk, T.: An application of the walkability index for elderly health—wieh. The case of the unesco historic centre of Porto, Portugal. Sustain. 13, (2021). https://doi.org/10.3390/su13094869.

86. Zuza Aranoa, M., Rivas Allo, C., Pérez-Ilzarbe Serrano, I.: Walkability City Tool (WCT): measuring walkability. Sci. Technol. 1, 13–15 (2016).

87. Grindlay, A.L., Ochoa-Covarrubias, G., Lizárraga, C.: Urban mobility and quality of public spaces: The case of granada, spain. In: WIT Transactions on the Built Environment. pp. 37–48 (2020). https://doi.org/10.2495/UT200041.

88. Boglietti, S., Tiboni, M.: Analyzing the Criticalities of Public Spaces to Promote Sustainable Mobility. In: Transportation Research Procedia. pp. 172–179. Elsevier B.V. (2022). https://doi.org/10.1016/j.trpro.2021.12.023.

89. Mastrolonardo, L.: Sustainable mobility and beauty of public space. Vitr. - Int. J. Archit. Technol. Sustain. 6, 42–55 (2021). https://doi.org/10.4995/vitruvio-ijats.2021.16560.

90. Rossetti, T., Lobel, H., Rocco, V., Hurtubia, R.: Explaining subjective perceptions of public spaces as a function of the built environment: A massive data approach. Landsc. Urban Plan. 181, 169–178 (2019). https://doi.org/10.1016/j.landurbplan.2018.09.020.

91. Karacor, E.K., Akcam, E.: Comparative Analysis of the Quality Perception in Public Spaces of Duzce City. Curr. Urban Stud. 04, 257–266 (2016). https://doi.org/10.4236/cus.2016.43017.

92. Ho, R., Au, W.T.: Scale Development for Environmental Perception of Public Space. Front. Psychol. 11, 596790 (2020). https://doi.org/10.3389/fpsyg.2020.596790.

93. Song, B., Park, K.: Contribution of greening and high-albedo coatings to improvements in the thermal environment in complex urban areas. Adv. Meteorol. 2015, 792172 (2015). https://doi.org/10.1155/2015/792172.

94. Delval, T., Geffroy, B., Rezoug, M., Jolibois, A., Oliveira, F., Carré, S., Tual, M., Soula, J.: BIM to Develop Integrated, Incremental and Multiscale Methods to Assess Comfort and Quality of Public Spaces. In: Lecture Notes in Civil Engineering. pp. 160–179 (2021). https://doi.org/10.1007/978-3-030-51295-8_14.

95. Silva, L.T., Mendes, J.F.G.: City Noise-Air: An environmental quality index for cities. Sustain. Cities Soc. 4, 1–11 (2012). https://doi.org/10.1016/j.scs.2012.03.001.

96. Costamagna, F., Lind, R., Stjernström, O.: Livability of Urban Public Spaces in Northern Swedish Cities: The Case of Umeå. Plan. Pract. Res. 34, 131–148 (2019). https://doi.org/10.1080/02697459.2018.1548215.

97. Güilden, D., Giritlioglu, C.: The evaluation of urban quality and vitality of the Istanbul historical peninsula- Eminönü district. Transportation (Amst). 5, 97–117 (2008).

98. Rebernik, N., Szajczyk, M., Bahillo, A., Marušic, B.G.: Measuring disability inclusion performance in cities using disability inclusion evaluation tool (DIETool). Sustain. 12, 1378 (2020). https://doi.org/10.3390/su12041378.

99. Dhasmana, P., Bansal, K., Kaur, M.: Assessing Gender Inclusive User Preferences: A case of Urban Public Spaces in Chandigarh. 2022 Int. Conf. Innov. Intell. Informatics, Comput. Technol. 3ICT 2022. 221–226 (2022). https://doi.org/10.1109/3ICT56508.2022.9990637.

100. Lopes, M., Santos Cruz, S., Pinho, P.: Using Publicness as a public space transdisciplinary analysis tool Analysis of urban metabolism in university campuses View project AESOP Public Spaces and Urban Cultures thematic group View project. In: Public Spaces – Urban Cultures Conference | FAUTL. , Lisbon (2012). https://doi.org/10.13140/RG.2.1.3874.3769.

101. Karaçor, E.K., Karacor, E.K.: Public vs. Private: The Evaluation of Different Space Types in Terms of Publicness Dimension. Eur. J. Sustain. Dev. 5, 51–58 (2015). https://doi.org/10.14207/ejsd.2016.v5n3p51.

102. Alwah, A., wen, I., Mofreh, B., Drmoush, A., Brook, K., Ali, N., Al-Shalafy, H., Alwah, M.: Intensity and Diversity of Use as a Tool to Measure the Quality of Public Spaces. J. Sustain. Cities Built Environ. 01, (2023). https://doi.org/10.58757/jscbe.idutmqps.03.

103. Németh, J., Schmidt, S.: Publicly Accessible Space and Quality of Life: A Tool for Measuring the Openness of Urban Spaces. In: Budruk, M. and Phillips, R. (eds.) Quality-of-Life Community Indicators for Parks, Recreation and Tourism Management. pp. 41–66. Springer Netherlands, Dordrecht (2011). https://doi.org/10.1007/978-90-481-9861-0_4.

104. Lee, H.-G., Lee, J.-H.: A Study on the Design Evaluation Indicators and Improvements of Urban Public Space. J. Korea Acad. Coop. Soc. 16, 8021–8029 (2015). https://doi.org/10.5762/kais.2015.16.11.8021.

105. Kim, D.-G., Moon, J.-M.: Assessment Indexes for Activation of Urban Public Space. J. Korean Hous. Assoc. 24, 109–119 (2013). https://doi.org/10.6107/jkha.2013.24.3.109.

106. Kaghouche, M., Benkechkache, I.: Assessment of the Quality of Public Spaces in the New City of Ali Mendjeli in Constantine (Algeria). Bull. Serbian Geogr. Soc. 103, 161–176 (2023). https://doi.org/10.2298/GSGD2302161K.

107. Galiano, A., Echarri, V.: Analysis of the quality of public urban space through a graphical analysis method. WIT Trans. Ecol. Environ. 191, 1623–1624 (2014). https://doi.org/10.2495/SC141382.

108. Varna, G., Tiesdell, S.: Assessing the publicness of public space: The Star Model of publicness. J. Urban Des. 15, 575–598 (2010). https://doi.org/10.1080/13574809.2010.502350.

109. Oppio, A., Bottero, M., Arcidiacono, A.: Assessing urban quality: a proposal for a MCDA evaluation framework. Ann. Oper. Res. 312, 1427–1444 (2022). https://doi.org/10.1007/s10479-017-2738-2.

110. Gawlak, A., Matuszewska, M., Ptak, A.: Inclusiveness of urban space and tools for the assessment of the quality of urban life—a critical approach. Int. J. Environ. Res. Public Health. 18, (2021). https://doi.org/10.3390/ijerph18094519.

111. Yunitsyna, A., Shtepani, E.: Investigating the socio-spatial relations of the built environment using the Space Syntax analysis – A case study of Tirana City. Cities. 133, (2023). https://doi.org/10.1016/j.cities.2022.104147.

112. Jenne, A.K., Tunçer, B., Beirão, J.N., Schmitt, G.: Stratification of Public Spaces based on Qualitative Attribute Measurement. In: Proceedings of the International Conference on Education and Research in Computer Aided Architectural Design in Europe. pp. 581–590 (2017). https://doi.org/10.52842/conf.ecaade.2017.2.581.

113. Palicki, S.: Multi-Criteria Assessment Of Public Space From The Social Perspective. Real Estate Manag. Valuat. 23, 24–34 (2015). https://doi.org/10.1515/remav-2015-0033.

114. Bambó Naya, R., de la Cal Nicolás, P., Díez Medina, C., Ezquerra, I., García-Pérez, S., Monclús, J.: Quality of public space and sustainable development goals: analysis of nine urban projects in Spanish cities. Front. Archit. Res. 12, 477–495 (2023). https://doi.org/10.1016/j.foar.2023.01.002.

115. Mantey, D., Kepkowicz, A.: Models of community-friendly recreational public space in warsaw suburbs. Methodological approach. Sustain. 12, (2020). https://doi.org/10.3390/SU12176764.

116. Fachrudin, H.T.: Urban quality measurement and it's influence to sense of place. IOP Conf. Ser. Earth Environ. Sci. 452, 012139 (2020). https://doi.org/10.1088/1755-1315/452/1/012139.

117. Oppio, A., Forestiero, L., Sciacchitano, L., Dell'Ovo, M.: How to assess urban quality: a spatial multicriteria decision analysis approach [Come valutare la qualità urbana: un approccio di analisi decisionale spaziale multi-criteriale per gli spazi aperti pubblici]. Valori e Valutazioni. 28, 21–30 (2021). https://doi.org/10.48264/vvsiev-20212803.

118. Hoffimann, E., Campelo, D., Hooper, P., Barros, H., Ribeiro, A.I.: Development of a smartphone app to evaluate the quality of public open space for physical activity. An instrument for health researchers and urban planners. Landsc. Urban Plan. 177, 191–195 (2018). https://doi.org/10.1016/j.landurbplan.2018.05.005.

119. Herthogs, P., Tunçer, B., Schläpfer, M., He, P.: A Weighted Graph Model to Estimate People's Presence in Public Space The Visit Potential Model. In: Proceedings of the International Conference on Education and Research in Computer Aided Architectural Design in Europe. pp. 611–620 (2018). https://doi.org/10.52842/conf.ecaade.2018.2.611.

120. Gómez de Salazar, N.N., Martín, M.C., Chamizo Nieto, F.J., Rosa-Jiménez, C., Muñoz, J.B.: Indicator System to Measure the Qualities of Urban Space Affecting Urban Safety and Coexistence. IOP Conf. Ser. Mater. Sci. Eng. 960, 042051 (2020). https://doi.org/10.1088/1757-899X/960/4/042051.

121. Portnov, B.A., Saad, R., Trop, T., Kliger, D., Svechkina, A.: Linking nighttime outdoor lighting attributes to pedestrians' feeling of safety: An interactive survey approach. PLoS One. 15, (2020). https://doi.org/10.1371/journal.pone.0242172.

122. Wu, Y., Wang, J., Lau, S.S.Y.S.S.Y., Lau, S.S.Y.S.S.Y., Miao, Y.: An Improved Publicness Assessment Tool Based on a Combined Spatial Model: Case Study of Guangzhou, China. Sustain. 14, 14711 (2022). https://doi.org/10.3390/su142214711.

123. Cao, Q., Yang, X., Li, S., Cai, W.: A Study on the Perception of Public Space in Displaced Relocation. Proc. 2021 Int. Conf. Cult. Des. Soc. Dev. (CDSD 2021). 634, (2022). https://doi.org/10.2991/assehr.k.220109.057.

124. Karuppannan, S., Sivam, A.: Comparative analysis of utilisation of open space at neighbourhood level in three Asian cities: Singapore, Delhi and Kuala Lumpur. Urban Des. Int. 18, 145–164 (2013). https://doi.org/10.1057/udi.2012.34.

125. Aliyas, Z., Gharaei, M.: Utilization and physical features of public open spaces in Bandar Abbas, Iran. IIOAB J. 7, 178–183 (2016).

126. Nitidara, N.P.A., Sarwono, J., Suprijanto, S., Soelami, F.X.N.: The multisensory interaction between auditory, visual, and thermal to the overall comfort in public open space: A

study in a tropical climate. Sustain. Cities Soc. 78, 103622 (2022). https://doi.org/10.1016/j.scs.2021.103622.

127. Yin, Y., Zhang, D., Zhen, M., Jing, W., Luo, W., Feng, W.: Combined effects of the thermal-acoustic environment on subjective evaluations in outdoor public spaces. Sustain. Cities Soc. 77, 103522 (2022). https://doi.org/10.1016/j.scs.2021.103522.

128. Alijani, S., Pourahmad, A., Hatami Nejad, H., Ziari, K., Sodoudi, S.: A new approach of urban livability in Tehran: Thermal comfort as a primitive indicator. Case study, district 22. Urban Clim. 33, 100656 (2020). https://doi.org/10.1016/j.uclim.2020.100656.

129. Ahmadi, S., Yeganeh, M., Motie, M.B., Gilandoust, A.: The role of neighborhood morphology in enhancing thermal comfort and resident's satisfaction. Energy Reports. 8, 9046–9056 (2022). https://doi.org/10.1016/j.egyr.2022.07.042.

130. Andrade, H., Alcoforado, M.J., Oliveira, S.: Perception of temperature and wind by users of public outdoor spaces: Relationships with weather parameters and personal characteristics. Int. J. Biometeorol. 55, 665–680 (2011). https://doi.org/10.1007/s00484-010-0379-0.

131. Li, R., Ou, D., Pan, S.: An improved service quality measurement model for soundscape assessment in urban public open spaces. Indoor Built Environ. 30, 985–997 (2021). https://doi.org/10.1177/1420326X20925527.

132. Lopes, J. V., Paio, A., Beirâo, J.N., Pinho, E.M., Nunes, L.: Multidimensional Analysis of Public Open Spaces: Urban Morphology, Parametric Modelling and Data Mining. Proc. Int. Conf. Educ. Res. Comput. Aided Archit. Des. Eur. 1, 351–360 (2015). https://doi.org/10.52842/conf.ecaade.2015.1.351.

133. Mbarep, D.P.P., Hasibuan, H.S., Moersidik, S.S.: The perception of thermal comfort felt by people in Kalijodo green open space. J. Pengelolaan Sumberd. Alam dan Lingkung. 11, 380–386 (2021). https://doi.org/10.29244/jpsl.11.3.380-386.

134. Ahirrao, P., Khan, S.: Assessing public open spaces: A case of city nagpur, india. Sustain. 13, 4997 (2021). https://doi.org/10.3390/su13094997.

135. Zavadskas, E.K., Bausys, R., Mazonaviciute, I.: Safety evaluation methodology of urban public parks by multi-criteria decision making. Landsc. Urban Plan. 189, 372–381 (2019). https://doi.org/10.1016/j.landurbplan.2019.05.014.

136. Kongphunphin, C., Srivanit, M.: Public spaces in Bangkok and the factors affecting the good public space quality in urban areas. IOP Conf. Ser. Mater. Sci. Eng. 910, 012024 (2020). https://doi.org/10.1088/1757-899X/910/1/012024.

137. Güngör, S., Polat, A.T.: The evaluation of the urban parks in Konya province in terms of quality, sufficiency, maintenance, and growth rate. Environ. Monit. Assess. 189, 172 (2017). https://doi.org/10.1007/s10661-017-5875-9.

138. Lorenzo, M., Ríos-Rodríguez, M.L., Suárez, E., Hernández, B., Rosales, C.: Quality analysis and categorisation of public space. Heliyon. 9, (2023). https://doi.org/10.1016/j.heliyon.2023.e13861.

139. Kraemer, R., Kabisch, N.: Parks in context: Advancing citywide spatial quality assessments of urban green spaces using fine-scaled indicators. Ecol. Soc. 26, (2021). https://doi.org/10.5751/ES-12485-260245.

140. Sağlik, A., Temiz, M., Kartal, F., Şenkuş, D.: Examining the Concept of Quality of Space in Public Open Spaces: The Example of Çanakkale Özgürlük Park. Mimar. Bilim. ve Uygulamaları Derg. 7, 795–812 (2022). https://doi.org/10.30785/mbud.1169558.

141. Harjanti, I.M.: Identification of Urban Park Quality in Taman Indonesia Kaya, Semarang. J. Archit. Des. Urban. 2, 1–14 (2020). https://doi.org/10.14710/jadu.v2i2.7001.

142. Malek, N.A., Mohammad, S.Z., Nashar, A.: Determinant factor for quality green open space assessment in Malaysia. J. Des. Built Environ. 18, 26–36 (2018). https://doi.org/10.22452/jdbe.vol18no2.3.

143. Zhaosen, Z., Guangsi, L.: Research on the Influencing Factors of Users' Perception of the Inclusiveness of Urban Parks Based on the Grounded Theory. Landsc. Archit. Front. 10, 12 (2022). https://doi.org/10.15302/j-laf-1-020066.

144. Nath, T.K., Zhe Han, S.S., Lechner, A.M.: Urban green space and well-being in Kuala Lumpur, Malaysia. Urban For. Urban Green. 36, 34–41 (2018). https://doi.org/10.1016/j.ufug.2018.09.013.

145. Ma, K.W., Mak, C.M., Wong, H.M.: Effects of environmental sound quality on soundscape preference in a public urban space. Appl. Acoust. 171, 107570 (2021). https://doi.org/10.1016/j.apacoust.2020.107570.

146. Bild, E., Pfeffer, K., Coler, M., Rubin, O., Bertolini, L.: Public space users' soundscape evaluations in relation to their activities. An Amsterdam-based study. Front. Psychol. 9, 366858 (2018). https://doi.org/10.3389/fpsyg.2018.01593.

147. Cohen, P., Potchter, O., Schnell, I.: A methodological approach to the environmental quantitative assessment of urban parks. Appl. Geogr. 48, 87–101 (2014). https://doi.org/10.1016/j.apgeog.2014.01.006.

148. Cengiz, C., Cengiz, B., Bekci, B.: Environmental quality analysis for sustainable urban public green spaces management in Bartın, Turkey. J. Food, Agric. Environ. 10, 938–946 (2012).

149. Patel, S.: Public Spaces for All : How "Public" are Public Spaces? Case of Ahmedabad city's Riverfront Parks. 106 (2016). https://doi.org/10.13140/RG.2.2.28917.99047.

150.	Ekawati, S.A., Ali, M., Trisutomo, S., Ghani, R.C.A.: The Study of Public Open Space effectiveness in Makassar Waterfront City using Good Public Space Index (GPSI). In: IOP Conference Series: Materials Science and Engineering. Institute of Physics Publishing (2020). https://doi.org/10.1088/1757-899X/875/1/012003.

151.	Akay, M., Okumuş, D.E., Gökçe, P., Terzi, F.: Re-coding The Characteristics of Public Spaces: The Case of İstanbul. Iconarp Int. J. Archit. Plan. 7, 487–512 (2019). https://doi.org/10.15320/iconarp.2019.95.

152.	Dovey, K., Pafka, E.: What is walkability? The urban DMA. Urban Stud. 57, 93–108 (2020). https://doi.org/10.1177/0042098018819727.

153.	Liu, M., Han, L., Xiong, S., Qing, L., Ji, H., Peng, Y.: Large-Scale Street Space Quality Evaluation Based on Deep Learning Over Street View Image. Lect. Notes Comput. Sci. (including Subser. Lect. Notes Artif. Intell. Lect. Notes Bioinformatics). 11902 LNCS, 690–701 (2019). https://doi.org/10.1007/978-3-030-34110-7_58.

154.	Li, J.: Evaluation and Improvement Strategy of Street Space Quality in Lujiazui Core Area of Shanghai Based on Multi-source Data Fusion. Proc. 57th ISOCARP World Plan. Congr. (2022). https://doi.org/10.47472/rgbayq2v.

155.	Painter, K.: The influence of street lighting improvements on crime, fear and pedestrian street use, after dark. Landsc. Urban Plan. 35, 193–201 (1996). https://doi.org/10.1016/0169-2046(96)00311-8.

156.	Sulaiman, N., Ayu Abdullah, Y., Hamdan, H.: Street as Public Space - Measuring Street Life of Kuala Lumpur. IOP Conf. Ser. Mater. Sci. Eng. 245, 082025 (2017). https://doi.org/10.1088/1757-899X/245/8/082025.

157.	Charkhchian, M., Daneshpour, S.A.: Interactions among different dimensions of a responsive public space: Case study in Iran. Rev. Urban Reg. Dev. Stud. 21, 14–36 (2009). https://doi.org/10.1111/j.1467-940X.2009.00157.x.

158.	Doğan, U.: A comparison of space quality in streets in the context of public open space design: The example of Izmir, Barcelona, and Liverpool. J. Urban Aff. 45, 1282–1315 (2023). https://doi.org/10.1080/07352166.2021.1919018.

159.	Fan, W.-Y.: Analysis on Traditional Commercial Street Quality with a Neighborhood Communication Perspective —— A Case of Rue de Marchienne, Belgium. Sci. J. Bus. Manag. 4, 181 (2016). https://doi.org/10.11648/j.sjbm.20160405.16.

160.	Garau, C., Annunziata, A., Yamu, C.: A walkability assessment tool coupling multi-criteria analysis and space syntax: the case study of Iglesias, Italy. Eur. Plan. Stud. 32, 211–233 (2024). https://doi.org/10.1080/09654313.2020.1761947.

161. Aşilioğlu, F., Çay, R.D.: Determination of quality criteria of urban pedestrian spaces. Archit. City Environ. 15, 1–25 (2020). https://doi.org/10.5821/ace.15.44.9297.

162. Gürer, N., Imren Güzel, B., Kavak, I.: Evaluation on Living Public Spaces and Their Qualities - Case Study from Ankara Konur, Karanfil and Yüksel Streets. IOP Conf. Ser. Mater. Sci. Eng. 245, 072038 (2017). https://doi.org/10.1088/1757-899X/245/7/072038.

163. Barreda Luna, A.A., Kuri, G.H., Rodríguez-Reséndiz, J., Zamora Antuñano, M.A., Altamirano Corro, J.A., Paredes-Garcia, W.J.: Public space accessibility and machine learning tools for street vending spatial categorization. J. Maps. 18, 43–52 (2022). https://doi.org/10.1080/17445647.2022.2035836.

164. Tang, J., Long, Y.: Measuring visual quality of street space and its temporal variation: Methodology and its application in the Hutong area in Beijing. Landsc. Urban Plan. 191, 103436 (2019). https://doi.org/10.1016/j.landurbplan.2018.09.015.

165. Ait Bihi Ouali, L., Laffitte, C., Graham, D.J.: Quantifying the Impact of Street Lighting and Walkpaths on Street Inclusiveness: The Case of Delhi. SSRN Electron. J. (2021). https://doi.org/10.2139/ssrn.3909297.

166. Alonso de Andrade, P., Berghauser Pont, M., Amorim, L.: Development of a Measure of Permeability between Private and Public Space. Urban Sci. 2, 87 (2018). https://doi.org/10.3390/urbansci2030087.

167. Rodrigues, D.S., Ramos, R.A.R., Mendes, J.F.G.: Multi-dimensional evaluation model of quality of life in campus. WSEAS Trans. Inf. Sci. Appl. 6, 1882–1892 (2009).

168. Hassanain, M.A., Sanni-Anibire, M.O., Mahmoud, A.S.: Design quality assessment of campus facilities through post occupancy evaluation. Int. J. Build. Pathol. Adapt. 41, 693–712 (2023). https://doi.org/10.1108/IJBPA-04-2021-0057.

169. Devitofrancesco, A., Belussi, L., Meroni, I., Scamoni, F.: Development of an Indoor Environmental Quality Assessment Tool for the Rating of Offices in Real Working Conditions. Sustainability. 11, 1645 (2019). https://doi.org/10.3390/su11061645.

170. FitzZaland, E., Wyatt, A.: Urban Design: An Investigation into the Visual, Perceptual, and Social Dimensions of Public Space. Focus (Madison). 2, (2005). https://doi.org/10.15368/focus.2005v2n1.9.

171. Sun, S., Yu, Y.: Dimension and formation of placeness of commercial public space in city center: A case study of Deji Plaza in Nanjing. Front. Archit. Res. 10, 229–239 (2021). https://doi.org/10.1016/j.foar.2020.08.001.

172. Vieira, M.C., Sperandei, S., Reis, A., da Silva, C.G.T.: An analysis of the suitability of public spaces to physical activity practice in Rio de Janeiro, Brazil. Prev. Med. (Baltim). 57, 198–200 (2013). https://doi.org/10.1016/j.ypmed.2013.05.023.

173. Khajehpour, E., Rasooli, D.: The concept of quality in public courtyards: Explanations and analyses case study: Mausoleum of Shah Ni'mat-Allah Vali. Sp. Ontol. Int. J. 9, 25–45 (2020).

174. Tutut Subadyo, A., Tutuko, P., Cahyani, S.D.: Assessment of Inclusive Historical Public Spaces in achieving preservation of such areas in Malang, Indonesia: Case study: Public spaces developed during the Dutch Colonial period. Int. Rev. Spat. Plan. Sustain. Dev. 6, 76–92 (2018). https://doi.org/10.14246/irspsd.6.4_76.

175. Satar, A.E., Asmal, I., Syarif, E.: The Effectiveness of Utilizing Non-Green Public Space in Untia Fishermen Settlement. EPI Int. J. Eng. 4, 140–148 (2021). https://doi.org/10.25042/10.25042/epi-ije.082021.06.

176. Stauskis, G.: Methodology for testing and evaluating accessability in public spaces. T. Plan. Archit. 29, 147–154 (2005).

177. Cozens, P.M., Saville, G., Hillier, D.: Crime prevention through environmental design (CPTED): A review and modern bibliography. Prop. Manag. 23, 328–356 (2005). https://doi.org/10.1108/02637470510631483/FULL/PDF.

178. Evans, G.W., McCoy, J.M.: WHEN BUILDINGS DON'T WORK: THE ROLE OF ARCHITECTURE IN HUMAN HEALTH. J. Environ. Psychol. 18, 85–94 (1998). https://doi.org/10.1006/JEVP.1998.0089.

179. Steinfeld, E., Maisel, J.: Universal design : creating inclusive environments. (2012).

180. Mehta, V., Bosson, J.K.: Third Places and the Social Life of Streets. Environ. Behav. 42, 779–805 (2010). https://doi.org/10.1177/0013916509344677.

181. Bolici, R., Gambaro, M.: Urban security for the quality of public space. Techne. 104–113 (2020). https://doi.org/10.13128/techne-7824.

182. (UN-Habitat), U.N.H.S.P.: Global Public Space Toolkit: From Global Principles to Local Policies and Practice. , Nairobi, Kenya (2016).

183. Woolley, H., Rose, S., Carmona, M., Freedman, J.: The Value of Publlic Space: How high quality parks and public spaces create economic, social and environmental value. , London, UK (2004).

184. Rudd, A.: City-Wide Public Space Strategies: A guidebook for city leaders advance review copy. , Nairobi, Kenya (2020).

185. Mutai, J.: City-Wide Public Space Assessment Toolkit. , Nairobi, Kenya (2020).

186. Jian, I.Y., Luo, J., Chan, E.H.W.: Spatial justice in public open space planning: Accessibility and inclusivity. Habitat Int. 97, 102122 (2020). https://doi.org/10.1016/j.habitatint.2020.102122.

187. UN Habitat: Urban green spaces: A brief for action. Reg. Off. Eur. 24 (2017).

188. Carmona, M.: Public places urban spaces: The dimensions of urban design. Public Places Urban Spaces Dimens. Urban Des. 1–672 (2021). https://doi.org/10.4324/9781315158457.

189. Garau, P.: Public Space: Think piece. Peer Learn. Exch. Public Sp. (2014).